\documentclass[twocolumn]{aa}
\usepackage{graphicx}
\usepackage{txfonts}
\usepackage{natbib}
\bibpunct{(}{)}{;}{a}{}{,} 
\newcommand{\ha}  {H$\alpha$}

\def\simless{\mathbin{\lower 3pt\hbox
     {$\rlap{\raise 5pt\hbox{$\char'074$}}\mathchar"7218$}}}   
\def\simmore{\mathbin{\lower 3pt\hbox
     {$\rlap{\raise 5pt\hbox{$\char'076$}}\mathchar"7218$}}}   

\def\rsun{~{\rm R}_\odot}

\begin{document}

   \title{Long-term variability of high-mass X-ray binaries. \\ I.
Photometry\footnote{Finding charts with the identification of the secondary
standard stars are also available in electronic form at the CDS through the
VizieR service http://vizier.u-strasbg.fr/viz-bin/VizieR}}

   \subtitle{}
  \author{
        P. Reig\inst{1,2}
        \and
        J. Fabregat\inst{3}
          }

\authorrunning{Reig et al.}
\titlerunning{Photometric variability of HMXB}

   \offprints{pau@physics.uoc.gr}

   \institute{IESL, Foundation for Research and Technology-Hellas, 71110, 
                Heraklion, Greece 
         \and Physics Department, University of Crete, 71003, 
                Heraklion, Greece 
                \email{pau@physics.uoc.gr}
         \and Observatorio Astron\'omico, Universidad de Valencia, 
         E-46100 Burjassot, Spain  
        }

   \date{Received ; accepted}

\abstract{We present photometric observations of the field around the 
optical counterparts of high-mass X-ray binaries. Our aim is to study the
long-term photometric variability in correlation with their X-ray
activity and derive a set of secondary standard stars that can be used for
time series analysis. We find that the donors in Be/X-ray binaries
exhibit larger amplitude changes in the magnitudes and colours than those 
hosting a supergiant companion. The amplitude of variability increases with
wavelength in Be/X-ray binaries and remains fairly constant in supergiant
systems. When time scales of years are considered, a good correlation
between the X-ray and optical variability is observed. The X-rays cease when optical brightness decreases. These results reflect the fact that the
circumstellar disk in Be/X-ray binaries is the main source of both optical
and X-ray variability. We also derive the colour excess, $E(B-V)$,
selecting data at times when the contribution of the circumstellar disk was
supposed to be at minimum, and we revisit the distance estimates. }


   \maketitle

\section{Introduction}

Neutron star high-mass X-ray binaries (HMXB) are accretion-powered
binary systems where the neutron star orbits an early-type (O or B)
companion. The luminosity class of the optical companion subdivides HMXBs
into Be/X-ray binaries (BeXB), when the optical star is a dwarf, subgiant,
or giant OBe star (luminosity class III, IV, or V), and supergiant X-ray
binaries (SXRBs), if they contain an evolved star with luminosity class
I-II star.  In SXRBs, the optical star emits a  substantial  stellar 
wind, removing  between  $10^{-6}-10^{-8}$  M$_{\odot}$  yr$^{-1}$  with a
terminal velocity up to 2000 km s$^{-1}$.  A neutron star in a relatively 
close orbit will capture a  significant  fraction of this wind, sufficient
to power a bright X-ray source. In BeXB, the donor is a Be star, i.e. a
rapidly rotating star of spectral type early B or late O with a gaseous
circumstellar envelope. In classical Be stars, the circumstellar matter is
distributed in a dust-free, ionized Keplerian disk around the star's
equator \citep{rivinius13a}. In BeXB, the main source of matter available
for accretion is this circumstellar disk. 

HMXBs represent important laboratories to probe physical processes in
extreme conditions of gravity and magnetic fields and to study a large
number of fundamental astrophysical questions, such as star formation
\citep{grimm03,mineo12},  masses of neutron stars and their equation of
state \citep{meer07,tomsick10,manousakis12}, wind physics
\citep{negueruela10} and particle acceleration, including  supersonic and
relativistic fluid dynamics, emission mechanisms, and radiation
reprocessing\citep{hadrava12,bosch13}.

We have been monitoring the HMXBs visible from the Northern Hemisphere in
the optical band since 1998. The monitoring consists of $BVRI$ photometry
and medium resolution spectra around the \ha\ line. Since 2013, we also
obtain polarimetry in the $R_c$ band. Here we focus on the photometry and
present the results of more than ten years worth of  observations. In
addition to the targets, we obtained standard $BVRI$ photometry of a large
sample of stars in the HMXBs fields, from which we defined a list of
secondary standard stars in the Johnson-Cousins system in each field.
Secondary standards will help the monitoring of these objects in the
future, as they will allow the obtention of standard photometry without the
need of setting up all-sky photometric observing campaigns.

\begin{table*}
\caption{List of targets.}
\label{list}
\begin{center}
\begin{tabular}{llcclccll}
\hline  \hline
X-ray           &Optical        &RA         &DEC        &Spectral       &$E(B-V)$       &Distance       &disk-  &X-ray \\
name            &name           &(eq. 2000) &(eq. 2000) &type           &(mag.)         &(kpc)          &loss   &outburst  \\
\hline
2S\,0114+65     &V662 Cas       &01 18 02.7 &+65 17 30  &B1Ia           &1.33$\pm$0.04    &5.9$\pm$1.4 &--     &--    \\
4U\,0115+63     &V635 Cas       &01 18 31.8 &+63 44 33  &B0.2Ve         &1.71$\pm$0.05    &6.0$\pm$1.5 &yes    &II    \\
IGR\,J01363+6610&--             &01 35 49.5 &+66 12 43  &B1Ve           &1.61$\pm$0.03    &2.2$\pm$0.5         &no     &--    \\
IGR\,J01583+6713&--             &01 58 18.4 &+67 13 23  &B2IVe          &1.44$\pm$0.04    &3.4$\pm$0.8 &no     &--    \\
RX\,J0146.9+6121&LS I +61 235   &01 47 00.2 &+61 21 24  &B1Ve           &0.88$\pm$0.03    &2.5$\pm$0.6 &no     &--    \\
RX\,J0240.4+6112&LS I +61 303   &02 40 31.7 &+61 13 46  &B0.5Ve         &1.09$\pm$0.03    &1.6$\pm$0.4 &no     &--    \\ 
V\,0332+53      &BQ Cam         &03 34 59.9 &+53 10 23  &O8.5Ve         &1.94$\pm$0.03    &6.0$\pm$1.5 &no     &II    \\
RX\,J0440.9+4431&LS V +44 17    &04 40 59.3 &+44 31 49  &B0.2Ve         &0.91$\pm$0.03    &2.2$\pm$0.5         &yes    &I     \\
1A\,0535+262    &V725 Tau       &05 38 54.6 &+26 18 57  &O9.7IIIe       &0.77$\pm$0.04    &2.1$\pm$0.5 &no     &I,II  \\
IGR\,J06074+2205&--             &06 07 26.6 &+22 05 48  &B0.5Ve         &0.86$\pm$0.03    &4.1$\pm$1.0 &yes    &--    \\
AX\,J1845.0-0433&--             &18 45 01.5 &--04 33 58 &O9Ia           &2.42$\pm$0.07    &5.5$\pm$1.5 &--     &flare \\
4U\,1907+09     &--             &19 09 37.9 &+09 49 49  &O9.5Iab        &3.31$\pm$0.10    &4.4$\pm$1.2 &--     &      \\
XTE\,J1946+274  &--             &19 45 39.4 &+27 21 56  &B0--1IV-Ve     &1.18$\pm$0.04    &7.0$\pm$2.0 &no     &I,II  \\
KS\,1947+300    &--             &19 49 35.5 &+30 12 32  &B0Ve           &2.01$\pm$0.05    &8.0$\pm$2.0 &no     &I,II  \\
EXO\,2030+375   &--             &20 32 15.3 &+37 38 15  &B0Ve           &3.00$\pm$0.20    &6.5$\pm$2.5   &no     &I,II  \\
GRO\,J2058+42   &--             &20 58 47.5 &+41 46 37  &O9.5--B0IV-Ve  &1.37$\pm$0.03    &9.0$\pm$2.5 &no     &--    \\
SAX\,J2103.5+4545&--            &21 03 35.7 &+45 45 06  &B0Ve           &1.36$\pm$0.03    &6.0$\pm$1.5 &yes    &I     \\
IGR\,J21343+4738&--             &21 34 20.4 &+47 38 00  &B1IV shell     &0.75$\pm$0.03    &10.0$\pm$2.5 &yes     &--    \\
4U\,2206+54     &BD+53 2790     &22 07 56.2 &+54 31 06  &O9.5Ve         &0.51$\pm$0.03    &3.0$\pm$0.7 &--     &--    \\
SAX\,J2239.3+6116&--            &22 39 20.9 &+61 16 27  &B0--2III-Ve    &1.66$\pm$0.04    &4.1$\pm$1.3 &no     &--    \\
\hline
\end{tabular}
\end{center}
\end{table*}

\begin{table*}
\caption{Average magnitudes and amplitude of variability of the targets.
$D_j$ is the difference between the largest and smallest measured
magnitudes.}
\label{target}
\begin{center}
\begin{tabular}{lcccccccccccccc}
\hline  \hline
Source          &$\overline{B}$   &$\overline{V}$   &$\overline{R}$   &$\overline{I}$    &$D_B$ &$D_V$ &$D_R$ &$D_I$  &$N_B$  &$N_V$  &$N_R$  &$N_I$  \\
                &(mag) &(mag) &(mag) &(mag)  &(mag) &(mag) &(mag) &(mag)  &        &       &       &       \\     
\hline
2S\,0114+65     &12.17 &11.03 &10.33 &9.58   &0.17  &0.15  &0.16  &0.09   &10 &10 &10 &8  \\
4U\,0115+63     &16.92 &15.34 &14.34 &13.22  &0.67  &0.85  &1.02  &1.04   &27 &27 &27 &22 \\
IGR\,J01363+6610&14.72 &13.31 &12.32 &11.37  &0.21  &0.16  &0.16  &0.17   &16 &16 &16 &16 \\
IGR\,J01583+6713&15.71 &14.41 &13.51 &12.66  &0.07  &0.08  &0.08  &0.06   &9  &9  &9  &9  \\
RX\,J0146.9+6121&12.09 &11.42 &11.00 &10.52  &0.20  &0.20  &0.18  &0.19   &26 &26 &26 &20 \\
RX\,J0240.4+6112&11.61 &10.75 &10.19 &9.55   &0.19  &0.15  &0.12  &0.16   &11 &11 &10 &10 \\
V\,0332+53      &17.16 &15.42 &14.26 &13.04  &0.28  &0.28  &0.30  &0.35   &9  &9  &9  &7  \\
RX\,J0440.9+4431&11.42 &10.73 &10.28 &9.76   &0.18  &0.34  &0.42  &0.48   &14 &14 &14 &14 \\
1A\,0535+262    &9.74  &9.19  &8.77  &8.30   &0.27  &0.32  &0.45  &0.44   &7  &7  &7  &6  \\
IGR\,J06074+2205&12.85 &12.21 &11.80 &11.32  &0.12  &0.27  &0.35  &0.45   &6  &6  &6  &6  \\
AX\,J1845.0-0433&16.24 &14.06 &12.71 &11.42  &0.19  &0.11  &0.08  &0.06   &11 &11 &11 &9  \\ 
4U\,1907+09     &19.41 &16.35 &14.40 &12.53  &0.20  &0.18  &0.10  &0.08   &14 &14 &14 &10 \\
XTE\,J1946+274  &18.76 &16.92 &15.62 &14.38  &0.31  &0.16  &0.14  &0.11   &7  &7  &7  &6  \\
KS\,1947+300    &15.16 &14.22 &13.53 &12.88  &0.14  &0.10  &0.13  &0.13   &18 &18 &18 &15 \\
EXO\,2030+375   &22.16 &19.41 &17.32 &15.18  &0.6\tablefootmark{$\dag$}  &0.23  &0.14  &0.18   &6  &8  &8  &8  \\
GRO\,J2058+42   &16.04 &14.89 &14.16 &13.35  &0.21  &0.25  &0.35  &0.44   &21 &21 &21 &17 \\
SAX\,J2103.5+4545&15.34&14.20 &13.49 &12.75  &0.39  &0.58  &0.70  &0.86   &21 &21 &21 &18 \\
IGR\,J21343+4738&14.68 &14.16 &13.80 &13.42  &0.23  &0.18  &0.17  &0.12   &6  &6  &6  &6  \\
4U\,2206+54     &10.11 &9.84  &9.64  &9.43   &0.19  &0.20  &0.09  &0.13   &27 &28 &27 &23 \\
SAX\,J2239.3+6116&16.26 &14.80 &13.89 &12.92 &0.15  &0.21  &0.28  &0.28   &7  &7  &7  &7  \\
\hline
\end{tabular}
\tablefoot{
\tablefoottext{$\dag$}{Affected by a large uncertainty due to the faintness of the source in this band.} 
}
\end{center}
\end{table*}

\section{Observations}

The data presented in this work were obtained from the Skinakas
Observatory, located in the island of Crete (Greece).  Therefore, the list
of targets given in Table~\ref{list} includes HMXBs that are visible from
the Northern Hemisphere (source declination $\simmore-20^\circ$). The
observations cover the period 2001-2013. We analysed a total of 45 nights. The instrumental set-up consisted of the 1.3 m telescope, the
Johnson-Cousins-Bessel  ($B$, $V$, $R$, and $I$ filters) photometric system
\citep{bessel90} and a CCD camera. We employed two different CCD chips.
Before 2007 June, a 1024$\times$ 1024 SITe chip with a 24 $\mu$m pixel size
(corresponding to 0.5 arcsec on the sky) was used. From 2007 July, the
telescope was equipped with a  2048$\times$2048 ANDOR CCD with a 13.5
$\mu$m pixel size (corresponding to 0.28 arcsec on the sky), and thus
providing a field of view of $\sim$9.5 arcmin squared. 

We carried out a reduction of the data in the standard way, using the IRAF
tools for aperture photometry. First, all images were bias-frame subtracted
and flat-field corrected using twilight sky flats to correct for
pixel-to-pixel variations on the chip. The resulting images are therefore
free from the instrumental effects.   We took the absorption caused by the Earth's
atmosphere  into account with nightly extinction corrections
determined from measurements of selected stars that also served as
standards. Finally, the photometry was accurately corrected for colour
equations and transformed to the standard system using nightly observations
of standard stars from Landolt's catalogue \citep{landolt92,landolt09}.
The linear transformation equations for each filter are of the form

\begin{equation}
\label{trans}
m^{0}_{j}= M_{j}+t_{j} \times SC + \kappa_{j}
\times X_{j} + Z_{j}
,\end{equation}

\noindent where $m^{0}_{j}$ is the atmospheric-extinction-corrected
instrumental magnitudes, $M_{j}$ the standard value, $\kappa_{j}$ is the
atmospheric absorption coefficient, $X_{j}$ is the airmass, $t_{j}$
a colour coefficient accounting for the differences in spectral response,
SC is a standard star colour, and $Z_{j}$ is the zero point of the telescope. The
subindex $j$ runs over different filters. We used $SC=(B-V)$ as the colour
term for the $B$ and $V$ filters and $SC=(V-R)$ and $SC=(V-I)$ for the $R$
and $I$ filters, respectively.

We calculated the error of the photometry in the individual nights as the
standard deviation of the difference between the derived final calibrated
magnitudes of the standard stars and the magnitudes of the catalogue.
Typically errors are of the order of 0.01--0.03 magnitudes.

In general, all the light inside an aperture
with radius equal to 16 pixels was summed up to produce the instrumental
magnitudes. Because of the presence of near-by objects and larger pixel size of
the older CCD, the aperture radius was reduced for XTE\,J1946+27 and
4U\,1907+09. In the case of XTE\,J1946+27, the aperture radius was taken to
be 10 pixels when the ANDOR CCD was used and 8 pixels in observations
before 2007. For 4U\,1907+09, we used 10 pixels in the observations
obtained with the older CCD.  We determined the sky background as the
statistical mode of the counts inside an annulus 5 pixels wide and 20
pixels from the center of the object.

\section{Data analysis and selection criteria for secondary standards}
\label{crit}

Although differential photometry using nearby stars has been performed on
some of the targets before
\citep{mendelson91,bell93,finley94,goranskii01,larionov01,baykal05,kiziloglu07b,sarty09,kiziloglu09},
this is the first systematic attempt to define secondary standard stars
within a few arcminutes of the targets that are suitable for variability
studies in a large sample of sources.

The use of comparison stars, i.e. constant stars in the vicinity of the
targets, permits differential photometry to be performed. Differential
photometry is most suitable to investigate optical photometric variability
on short (seconds to minutes) time scales. In addition, if these constant
stars have well-determined magnitudes and colours in the standard system,
they can be used as local standards to obtain standard photometry of the
targets. This approach has a number of advantages with respect to the
classical all-sky absolute photometry: it allows us to reach a higher
precision; it removes the need to observe standard fields all around the
sky, thus saving valuable observing time; it can be performed under less
stringent weather conditions; and it eliminates the influence of
instrumental drifts, since they affect every star equally.

To find constant stars in the field of view of HMXB, we first extracted the
coordinates of a large number of stars using the IRAF task {\tt DAOFIND}.
This task searches for  local  density  maxima, which  have a given
full-width half-maximum (FWHM) and a peak amplitude greater than  a given
threshold  above the  local  background. Once the FWHM was fixed, we
adjusted the threshold parameter to achieve at least $\sim$40 detections in
low-populated fields and $\sim$100 in high-populated fields. Because we
obtained the data using different instrumental set-up, the resulting frames
differed in orientation and size, making the identification of the same
star a time-consuming task. To solve this issue, we calculated the relative
position of each star with respect to the target. We converted these
relative positions  into absolute positions in each frame by adding the
coordinates of the target in that frame. Likewise, the image coordinates
were scaled down in the smaller size frames (i.e. observations before 2007)
by multiplying by appropriate correction factors. 

Instrumental magnitudes were obtained for the target and each of the
detections and  then transform  to the standard system. The final
product was the calibrated $BVRI$ magnitudes as a function of time. 
Then we derived the average and standard deviation of the data


\begin{equation}
\label{mean}
\overline{m}_{j}=\frac{\sum\limits_{i=1}^{N_j}{m_{i,j}}}{N_j} 
\end{equation}
\begin{equation}
\label{stdev}
\sigma{_j}=\sqrt{\frac{\sum_{i=1}^{N_j}{(m_{i,j}-\overline{m}_j)^2}}{N_j-1}}
.\end{equation}

\noindent The index $j$ represents one of the four filters ($BVRI$) and $i$
runs over the number of photometric measurements.The quantities $N_j$ are
the total number of good measurements in each filter. In addition, we also
calculated the standard deviation of the mean as

\begin{equation}
\label{emean}
\sigma_{\overline{m_j}}=\frac{\sigma_j}{\sqrt{N_j}}
.\end{equation}

Secondary standard stars were selected using the following criteria:

\begin{itemize} \item[-] Variability. Variable stars can be identified as
those for which the standard deviation of the measured standard magnitudes
is significantly larger than the precision of the standard photometry on
individual nights. The standard deviation of the standard photometry is
between 0.01 and 0.03 magnitudes for most of the observing nights, as
stated above. In addition, measurements of individual stars are also
affected by instrumental photometry errors, and by higher photon noise if
the star is fainter than the standard stars used to calculate the
transformation coefficients. Taking this into account, we consider as
non-variable stars those for which the standard deviation (Eq.~\ref{stdev})
of the measurements in BVR is less or equal than 0.05 magnitudes. Because
the sky is significantly brighter in the $I$ band, the final magnitudes
resulted in slightly larger errors. Thus for the $I$ band we allowed a
larger value of 0.08 magnitudes.

\item[-] Number of observations. We rejected stars with fewer than six
measurements.

\item[-] Colour. The $(B-V)$ colour index of the comparison stars brackets
that of the target object. In most cases, the maximum allowed difference in
$(B-V)$ between the target and the comparison star is 0.5 magnitudes.  The
only exceptions are 4U\,1909+07 and EXO\,2030+375, which because of the very
high extinction the comparison stars are all bluer than the target.

\end{itemize}

We imposed no limit on the magnitude of the star. Ideally, the comparison
stars should be brighter than the target so that the photon noise of the
target dominates that of the standards. This condition is difficult to
meet if the target is bright and the field is not very populated. However,
for bright objects it does not represent a problem because  the secondary
standards, although fainter than the target, are bright enough and are
photon-noise dominated.  For relatively faint targets ($V_{\rm
src} \simmore14$ mag), the selection criteria generally pick up stars that
do not generally differ by more than 1 magnitude from the target value.

The results of our analysis of the field stars are presented in the
Appendix. We list a number of stars in the vicinity of the targets that have
remained constant over year's time scales. Thus these stars
represent good secondary standard stars suitable for calibration purposes
and variability studies. 

\begin{figure}
\begin{center}
\includegraphics[width=8cm,trim=40 20 60 40]{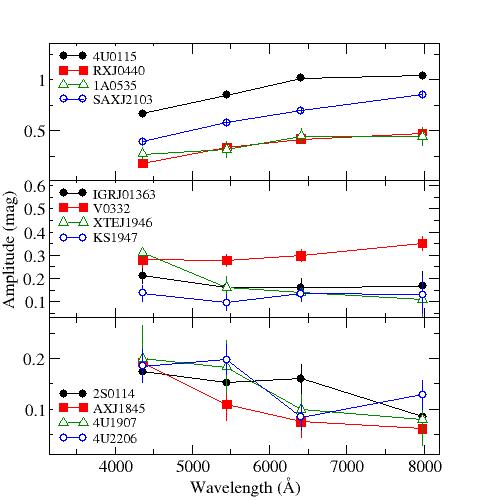}
\end{center}
\caption[]{Amplitude of magnitude change as a function of wavelength for
BeXB that have gone through disk-loss episodes (upper panel), BeXB with 
stable disks (middle panel), and SGXB and other wind-fed systems (lower panel). }
\label{excess}
\end{figure}

\section{Results on high-mass X-ray binaries}

In this section we present the results of our photometric analysis of the
targets.  Table~\ref{target} lists the results of the observations. In this
table, column 1 is the name of the X-ray source. Columns 2--5 contain the
mean magnitudes obtained from eq.(\ref{mean}). Columns 6--9 give the difference in magnitude between the
maximum and minimum values. Finally, columns 10--13 indicate the number of
observations used to derive the mean and the dispersion.

\begin{figure}
\begin{center}
\includegraphics[width=8cm,trim=60 40 180 80]{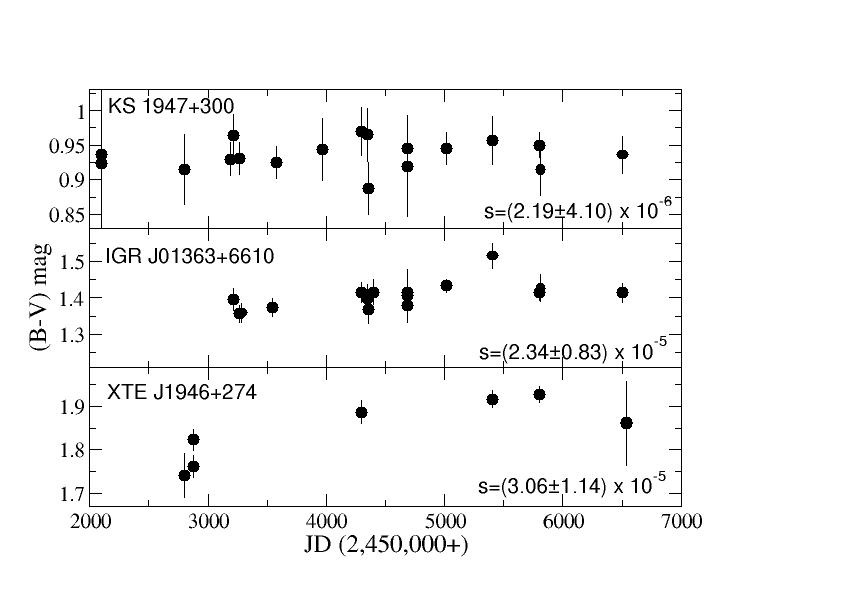}
\end{center}
\caption[]{Evolution of the $(B-V)$ colour index in three BeXB with stable
and lasting disks. $s$ is the slope to a linear fit.}
\label{bv}
\end{figure}

\subsection{Photometric variability}

Be stars are known to display photometric variability on all time scales.
While rapid variability is associated with the star's rotation and non-radial
pulsations \citep{balona11,kiziloglu07b,sarty09,gutierrez-soto11},
long-term variability is attributed to structural changes in the
circumstellar disk. In BeXBs, the largest amplitude variations are observed
on time scales of years \citep{lyuty00,reig07b}, which represent 
the typical time scale for formation and dissipation of the disk
\citep{jones08}.

Column 8 in Table~\ref{list} indicates whether the systems lost the disk
during the interval covered by our observations\footnote{1A\,0535+262 wend
through a disk-loss episode in 1998 \citep{haigh99}.}. By disk-loss episode, we mean that the equivalent width of the \ha\ line was measured
positive, indicating an absorption dominated line. It is possible that a
weak and small disk may be present even when emission is not detected
\citep[see e.g][]{reig14a}, but it definitively implies a low optical state.

\begin{figure}
\begin{center}
\includegraphics[width=8cm,trim=50 40 180 30]{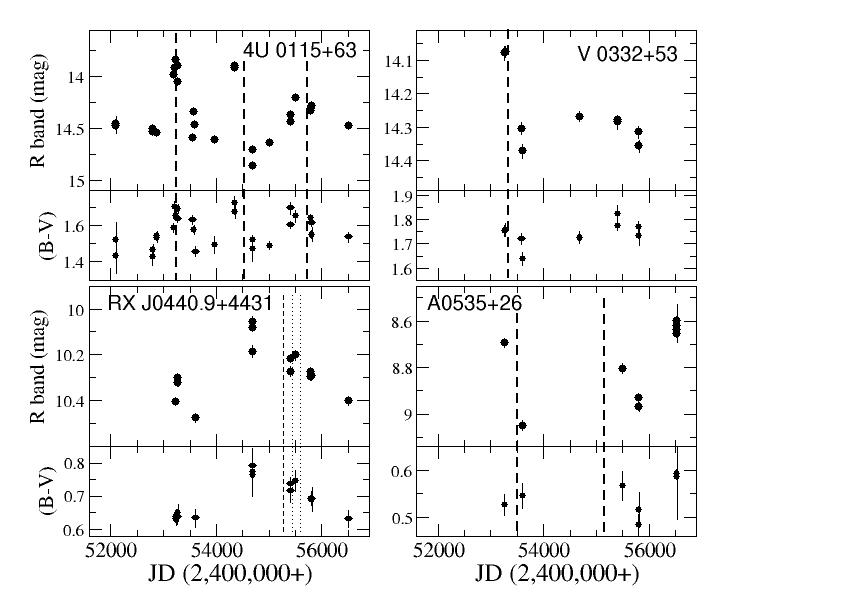}
\end{center}
\caption{$R$-band and $(B-V)$ light curves of four BeXB displaying large
amplitude X-ray activity. The dashed and dotted vertical lines indicate 
the epochs when type II and type I X-ray outbursts occurred, respectively.  
A decrease in the optical brightness following the outburst 
is seen in all systems.}
\label{xopt}
\end{figure}

In general, systems containing Be stars display much larger amplitude of
variability in the photometric magnitudes and colours than those harbouring
a supergiant companion. In the blue part of the spectrum, where $\Delta B\simless 0.2$ mag, SGXRBs tend to be more variable compared to $\Delta I <
0.1$ mag. In contrast, BeXBs, especially those systems  that have gone
through episodes of disk dissipation and reformation, such as 4U\,0115+63,
1A\,0535+262,  RX J0440.9+4431, and SAX\,J2103.5+4545  show the largest
magnitude difference in the redder bands with $\Delta I >> 0.1$ mag. 
Note however, that there are BeXBs with stable disks. These systems may
show moderate changes in the individual photometric bands, but the colours
hardly change on long time scales. These results are illustrated in
Figs.~\ref{excess} and \ref{bv}. Fig.~\ref{excess} displays the amplitude
of variability as a function of the effective wavelength for the four
bandpasses considered for four BeXBs that have gone through disk-loss
episodes (upper panel), four BeXB with stable disks (middle panel),  four
wind-fed systems, three SGXBs, and 4U\,2206+54 (lower panel). These three
groups of systems exhibit different behaviour. The amplitude of variability
in BeXB that have gone through disk-loss episodes increases with
wavelength, whereas that in wind-fed systems it decreases. The BeXB with
long-lived disks show no obvious trend with wavelength. Fig~\ref{bv}
displays the evolution of the $(B-V)$ colour index for three BeXB with
stable disks. A linear fit to the data shows that the slope is consistent
with zero. It is interesting to compare this figure with Fig.~\ref{xopt},
where the colour index of two optically variable BeXB is shown and where
large amplitude changes are apparent.

The reason for this different behaviour is that while in BeXBs the
variability comes from the disk, in supergiant stars it simply reflects the
varying but relatively constant (on long time scales) mass-loss rate from
the massive companion. The contribution of the disk to the overall
continuum emission increases with wavelength because the slope of the
thermal emission from the photosphere falls much faster than the spectral
energy distribution of the free-free radiation (electron bremsstrahlung)
from the disk at increasing wavelengths \citep{gehrz74}.

\begin{figure}
\begin{center}
\includegraphics[width=8cm,trim=60 40 180 30]{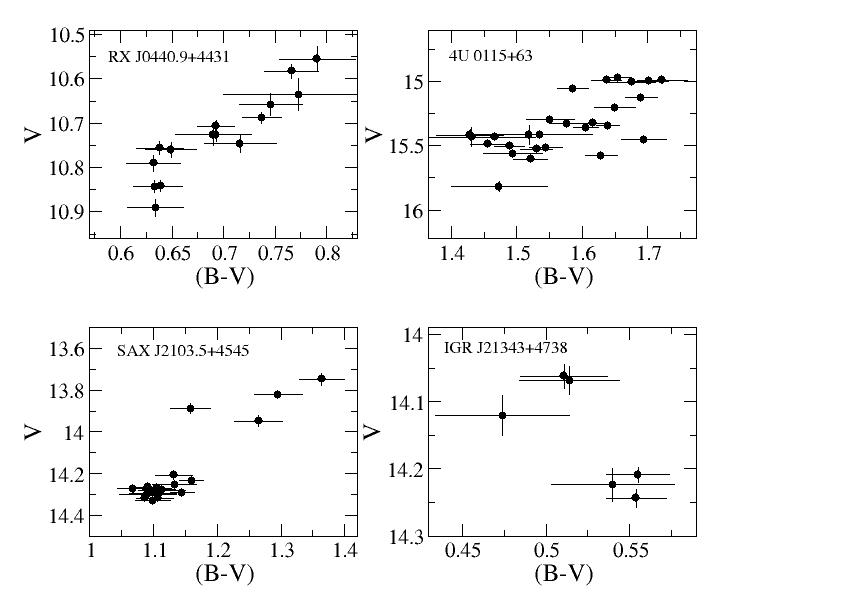}
\end{center}
\caption[]{$(B-V)-V$ colour magnitude diagram for four BeXB that have gone
through disk-loss. Note the different behaviour of IGR\,J21343+4738, the
only BeXB with a Be shell companion.}
\label{bvv}
\end{figure}

Two types of long-term correlated optical photometric and
spectroscopic variations have been observed in classical Be stars
\citep{harmanec00,harmanec83}. A positive correlation is characterised by
an increase in the brightness of the star as the strength of the H I lines
increases. In addition, the initial brightening in the optical (e.g. $V$
band) is accompanied by an increase in the $(B-V)$ colour. In other words,
as the disk reforms after a disk-loss phase, the systems becomes brighter
and redder. On the other hand, fewer stars show a negative correlation,
that is,  the stronger the \ha\ emission, the fainter the star. In this
case, the initial fading in the $V$ band is also accompanied by a reddening
of $(B-V)$. These correlations are attributed to a geometrical effect
\citep{harmanec83}.  Stars viewed at very high inclination angles show the
inverse correlation, whereas stars seen at certain inclination angles $i <
i_{\rm crit}$, exhibit the positive correlation. At very high inclination
angles (equator-on stars), the inner parts of the Be envelope partly block
the stellar photosphere, while the small projected area of the disk on the
sky keeps the disk emission to a minimum. In stars not seen equator-on, the
effect of the disk is to increase the effective radius of the star: as the
disk grows an overall (star plus disk) increase in brightness is expected.
The value of the critical inclination angle is not known, but a rough
estimate based on available data suggest $i_{\rm crit}\sim 75^{\circ}$
\citep{sigut13}.

Figure~\ref{bvv}  shows four BeXB that have gone through a disk-loss phase.
The Be companion in IGR\,J21343+4738 is the only shell star of the four
\citep{reig14a}.  The presence of shell absorption lines indicate that the
line of sight to the star lies nearly perpendicular to its rotation axis.
While most of the BeXB show the positive correlation between $V$ and
$(B-V)$,  IGR\,J21343+4738 shows the negative correlation, hence confirming
\citet{harmanec83} explanation. The inverse correlation has been questioned
by \citet{haubois12} who argued that the changes of $(B-V)$ in shell stars
should be of rather small amplitude. Large change of the $(B-V)$ colour in
this case might be due to other causes, such as $V/R$ variations, rather
than to mass injection. The current data on IGR\,J21343+4738 show indeed a
much smaller change in $(B-V)$ with respect to the other three sources.
However, this may simply because of the fact that the disk growth is still in
the initial phase in this system as indicated by the weak \ha\ emission
\citep{reig14a}.

\subsection{Colour excess and distance}

In Be stars,  the total measured reddening is made up of two components:
one produced mainly by dust in the interstellar space through the line of
sight and another produced by the circumstellar gas around the Be star
\citep{dachs88,fabregat90}. Although the  physical origin and wavelength
dependence of these two reddenings is completely different, their final
effect upon the colours is very difficult to disentangle
\citep{torrejon07}. Indeed, interstellar reddening is caused by
"absorption" and "scattering" processes, while circumstellar reddening is
due to extra "emission" at longer wavelengths. In this respect, disk-loss
episodes are extremely important because they allow us to derive the true
magnitudes and colours of the underlying Be star, without the contribution
of the disk. 

The colour excess is simply $E(B-V)=(B-V)_{\rm obs}-(B-V)_0$, where
$(B-V)_{\rm obs}$ is the observed colour and $(B-V)_0$ the intrinsic colour. 
To minimise the contribution of the disk, we used the data with the bluest
colour index of each target.  The colour excess of the sources considered
in this work is listed in Table~\ref{list}. 

The uncertainty in estimating the distance stems from the uncertainty of
the spectral classification and of the intrinsic colour and absolute
magnitude calibrations. In particular, the error associated with the
absolute magnitude calibration can be large \citep{jaschek98,wegner06}. We
estimate the intrinsic colour and absolute magnitude from the spectral type
given in Table~\ref{list}. The adopted $(B-V)_0$ colour is the average of
the values from  \citet{johnson66}, \citet{fitzgerald70},
\citet{gutierrez-moreno79}, and \citet{wegner94} and the associated error,
the standard deviation.  The absolute magnitude was taken from
\citet{wegner06}, who gives different calibrations for emitting and
non-emitting B stars. For the error in $M_V$, we assumed 0.5 magnitudes. Only
for the systems where the spectral type is not well constrained we took 0.6
magnitudes.  The error in the distance was then obtained by propagating the
errors in $V$, $E(B-V)$, $(B-V)_0$, and $M_V$. Typically, the error in the
distance given in Table~\ref{list} is $\sim25$\%.

\begin{figure}
\begin{center}
\includegraphics[width=8cm,trim=40 70 180 80]{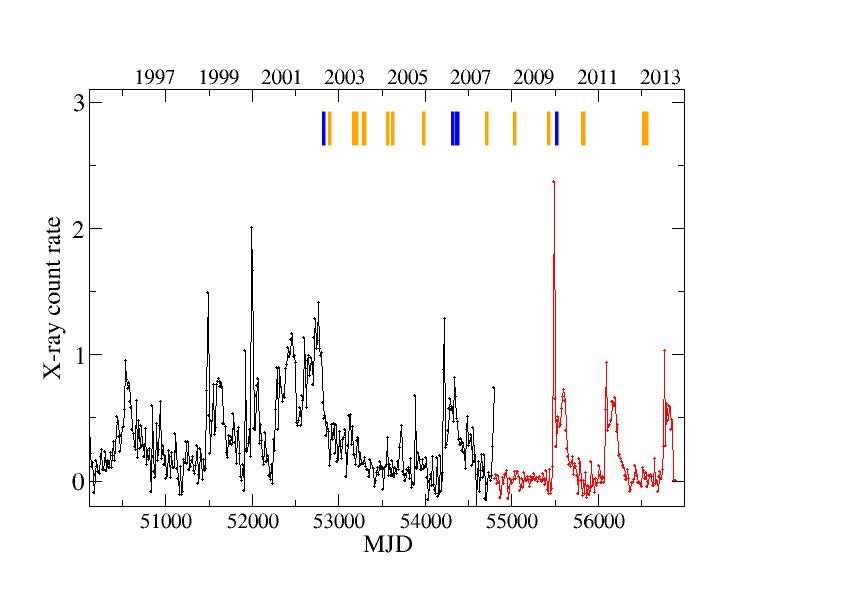}
\end{center}
\caption[]{Long-term X-ray variability of SAX\,J2103.5+4545 as seen by the 
all-sky monitors {\it RXTE}/ASM (black) and {\it SWIFT}/BAT light curves 
(red). The original one-day resolution light curves were rebinned to a bin 
size equal to the orbital period ($P_{\rm orb}=12.67$ d). The blue and
orange vertical lines indicate the epochs when the $BVRI$ magnitudes were 
below (brighter) and above (fainter) the average, respectively.
 }
\label{sax2103}
\end{figure}

\subsection{Correlated X-ray/optical variability}

Because the disk constitutes the main reservoir of matter available for
accretion, a correlation between X-rays and optical emission should be
expected. Although on short time scales (days-weeks, sometimes even months)
the relationship between optical emission and X-rays is complex and not
always seen, when one considers time scales of years, the correlation is
generally observed. The correlation identifies the circumstellar disk as
the  main source of optical {\em \emph{and}} X-ray variability. 

The material in the disk is the fuel that powers the X-rays through
accretion. The system shines bright in X-rays while there is enough fuel.
When this is exhausted the X-rays switch off. This is illustrated in 
Fig.~\ref{xopt}, where a significant decline in the optical brightness is
observed after the occurrence of X-ray outbursts.  Figure~\ref{sax2103}
shows the long-term X-ray light curve of SAX\,J2103.5+4545 using the
all-sky monitors on board {\it RXTE} and {\it Swift}\footnote{To bring the
BAT data points to the same scale as the ASM data points, we multiply the
BAT data by 129.3. This constant was derived by taking into account the
correspondence between intensity (mCrab) and count rate of the two
instruments (1 mCrab corresponds to 0.075 ASM count s$^{-1}$ and to 0.00022
BAT count cm$^{-2}$ s$^{-1}$) and the fact that the Crab emits about 38\%
more in the 2--10 keV than in the 15--50 keV band.}. The vertical blue and
orange lines correspond to observations in which the photometric magnitudes
were below (brighter) and above (fainter) the average, respectively (see
Table~\ref{target}). Clearly, low-intensity optical states are seen during
faint X-ray states, whereas during bright X-ray states the source displays
brighter optical magnitudes.

\begin{figure}
\begin{center}
\includegraphics[width=8cm,trim=50 60 60 230]{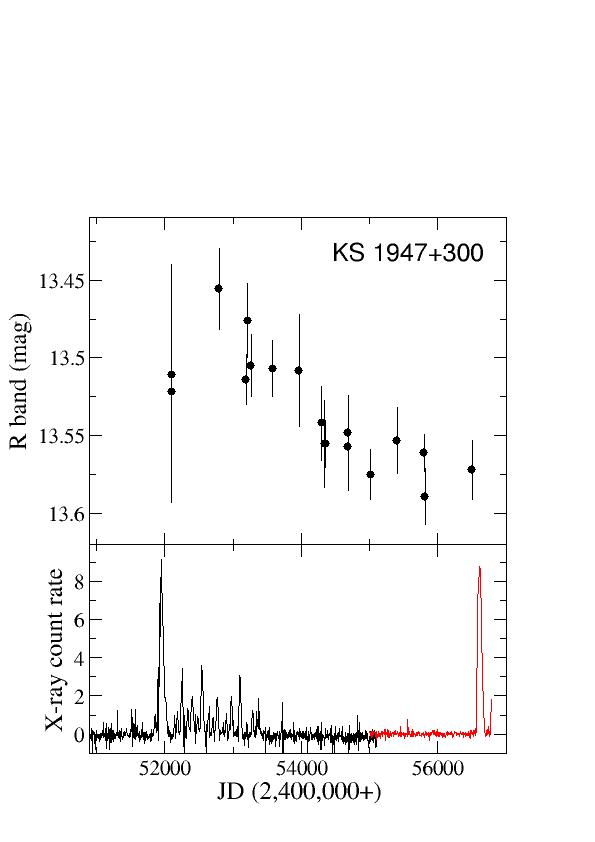}
\end{center}
\caption{$R$-band and X-ray light curves of KS\,1947+300 from the 
{\it RXTE}/ASM (black) and {\it Swift}/BAT (red) all-sky monitors. 
Note the longer disk variability time-scales (changes in the optical 
brightness) with respect to systems shown in Fig.~\ref{xopt}. }
\label{xopt_disk}
\end{figure}

Sometimes the source may be active in X-rays for a very long time even
after a major X-ray outburst, however,   the
X-rays cease once the matter supply ends. An example of this behaviour can be seen in
Fig.\ref{xopt_disk}, where the optical ($R$-band) and X-ray light curves
of  KS\,1947+300 are shown. The X-ray data come from the all-sky monitors
{\it RXTE}/ASM (2--10 keV) for observation before MJD 55000 and from {\it
Swift}/BAT (15--50 keV) after that date.  During the $\sim$ five years that
KS\,1947+300 was active in X-rays ($\sim$ MJD 51900--53500), the optical
brightness remained fairly constant at $R\sim 13.5$ magnitudes. After MJD
53500, $R$ decreased and the X-ray activity ceased. For the next $\sim$ eight
years the optical brightness did not vary significantly
(R=13.56$\pm$0.01) and the system remained in an off X-ray state.
Surprisingly, the source went into another outburst without a substantial
rebrightening in the optical magnitudes.

Historically, the long-term X-ray variability of BeXB has been
characterised by two type of outbursts.

\begin{itemize}

\item Type I outbursts are regular and (quasi-)periodic outbursts, normally
peaking at or close to periastron passage of the neutron star. They are
short-lived, i.e. tend to cover a relatively small fraction of the orbital
period (typically 0.2--0.3 $P_{\rm orb}$). The X-ray flux increases by up
to two orders of magnitude with respect to the pre-outburst state, reaching
peak luminosities $L_X \simless 10^{37}$ erg s$^{-1}$.  The number of
detected type I outbursts varies from source to source, but typically the
source remains active for 3--8 periastron passages. A remarkable case is
EXO 2030+375, which shows type I outbursts almost permanently.

\item Type II outbursts represent major increases of the X-ray flux,
$10^3-10^4$ times that at quiescence. They may reach the Eddington
luminosity for a neutron star ($L_X \sim 10^{38}$ erg s$^{-1}$) and become
the brightest objects of the X-ray sky. They do not show any preferred
orbital phase and last for a large fraction of an orbital period or even
for several orbital periods. 

\end{itemize}

Column 9 in Table~\ref{list} indicates the type of outburst detected during
the course of our observation in the interval 2001--2013. It should be
emphasized that as X-ray missions generate more data, it has become apparent
that classification based solely on the X-ray luminosity is too simplistic.
Outbursts in BeXB show a large variety of properties and many outbursts do
not always follow this simple scheme \citep{kretschmar12}.

In principle, a type II outbursts appears as a dramatic event that may lead
to the dissipation of the disk. In practice, the loss of the disk after a
major outburst is the exception rather than the rule. Of the systems that
exhibited type II outbursts, only 4U\,0115+63 loses the disk after major
outbursts \citep{reig07b}. Nevertheless, type II outbursts do affect the
structure of the disk (Fig.~\ref{xopt}) and fainter optical emission is
generally detected after such events.

Although estimating the total mass contained in a disk is rather uncertain
because of the uncertainty in its size and density, an order of magnitude
calculation of the amount of material that is accreted in  the two cases
would be illustrative. To generate an average $L_X \sim 1\times 10^{37}$
erg s$^{-1}$ over 100 days, a mass accretion rate of $\dot{M}\sim 2.7
\times 10^{17}$ g s$^{-1}$ is needed. Assuming that all the potential
energy loss in the accretion process is converted in radiation,
$\dot{M}=L_X(R/GM)$, then the total amount of matter accreted into the
neutron star is $4.7\times 10^{23}$ g. In the case of a type I outburst
with average luminosity, $L_X \sim 3\times 10^{35}$ erg s$^{-1}$, and a
duration of 20 days, the  total accreted mass would be  $2.8\times 10^{21}$
g.

The total mass of the disk can be computed as $M=\int{\rho\, dV}$, with
$\rho=\rho_0(R_*/r)^{n}$ and $dV=2\pi rH dr$. Here $\rho_0$ is the density
and the inner radius of the disk (at $r\sim R_*$), $R_*$ the star radius, 
$H$ the disk height, and $n$ an exponent defining the density law. The
parameter $n$ varies in the range 3--3.5 in most BeX \citep{waters88}. The
typical radius of a B0V star, is $R_*=8\rsun$ and assuming  typical values
for the disk radius $R_{\rm d}=5R_*$, and inner density $\rho_0=10^{-10}$
g  cm$^{-3}$ \citep{telting98} and $H=0.03R_*$ \citep{negueruela01} the
mass is estimated to be $2 \times 10^{24}$ g.

Therefore, while in a type II outburst the total accreted mass may
represent an important part of the disk, only an insignificant 0.1\% of the
total mass is accreted onto the neutron star during the course of a type I
outburst. Even if the outbursts are repeated several times, the disk may
not be substantially affected.

\section{Summary and conclusion}

We have studied the long-term optical photometric variability of the
counterparts to high-mass X-ray binaries visible from the Northern
Hemisphere. Our results can be summarised as follows:

\begin{enumerate}

\item There is a complex connection between the optical and X-ray
variability in BeXB. However, when time-scales of years are considered,
there is a correlation between the long-term evolution of the system
optical brightness and the X-ray activity. After a major X-ray outburst the
optical magnitudes are always fainter than before the outburst. 

\item The disk is the main source of optical variability because the disk
itself emits at these wavelength, but it is also responsible for the X-ray
variability because it constitutes the source of matter available for
accretion. As long as the disk is stable, no changes are seen in the colour
indices. In BeXB with stable disks, the amplitude change in magnitudes shows
almost no variation with wavelength.

\item The more variable the BeXB in the optical band, the more variable in
the X-ray band. Systems with the larger amplitude of variability in the
optical magnitudes and colours are those with also the larger amplitude
changes in the X-ray band.

\item The HMXB with Be companions display larger amplitude of optical
variability ($\Delta V >> 0.1$ mag) than SGXBs ($\Delta V \simless 0.1$ mag).

\item The amplitude of variability increases with wavelength in BeXB with
fast-changing disks and decreases in SGXBs. 

\item  The positive and negative (or inverse) correlations between
luminosity and colour during disk growth observed in classical Be stars are
also present in BeXBs. 

\item Contrary to general belief, type II outbursts do not generally lead
to the total destruction of the disk, although in some cases, such as
in 4U\,0115+63, it does lead to the total destruction of the disk. In this
system, the optical variability can be explained by a build up and
disruption of the circumstellar disk on times scales of three--five years.

\item We have also set up a list of secondary standard stars in the field
of view of high-mass X-ray binaries. This work can benefit observers who
seek to perform differential photometry for frequency analysis. It also
allows us to perform standard photometry without the framework of a full
all-sky absolute photometry observing campaign.

\end{enumerate}

\begin{acknowledgements}

The work of J.F. is supported by the Spanish Ministerio de Econom\'{\i}a y
Competitividad, and FEDER, under contract AYA2010-18352. P.R. and J.F. are
partially supported by the Generalitat Valenciana project of excellence
PROMETEOII/2014/060. Skinakas Observatory is a collaborative project of the
University of Crete and the Foundation for Research and Technology-Hellas. 
This work made use of NASA's Astrophysics Data System Bibliographic
Services and of the SIMBAD database, operated at the CDS, Strasbourg,
France. PR acknowledges partial help from the COST action MP1304
``Exploring fundamental physics with compact stars".

\end{acknowledgements}

\bibliographystyle{aa}

\begin{thebibliography}{47}
\expandafter\ifx\csname natexlab\endcsname\relax\def\natexlab#1{#1}\fi

\bibitem[{{Balona} {et~al.}(2011){Balona}, {Pigulski}, {Cat}, {Handler},
  {Guti{\'e}rrez-Soto}, {Engelbrecht}, {Frescura}, {Briquet}, {Cuypers},
  {Daszy{\'n}ska-Daszkiewicz}, {Degroote}, {Dukes}, {Garcia}, {Green}, {Heber},
  {Kawaler}, {Lehmann}, {Leroy}, {Molenda-{\.Z}aaowicz}, {Neiner}, {Noels},
  {Nuspl}, {{\O}stensen}, {Pricopi}, {Roxburgh}, {Salmon}, {Smith},
  {Su{\'a}rez}, {Suran}, {Szab{\'o}}, {Uytterhoeven}, {Christensen-Dalsgaard},
  {Kjeldsen}, {Caldwell}, {Girouard}, \& {Sanderfer}}]{balona11}
{Balona}, L.~A., {Pigulski}, A., {Cat}, P.~D., {et~al.} 2011, \mnras, 413, 2403

\bibitem[{{Baykal} {et~al.}(2005){Baykal}, {K{\i}z{\i}lo{\v g}lu},
  {K{\i}z{\i}lo{\v g}lu}, {Balman}, \& {Inam}}]{baykal05}
{Baykal}, A., {K{\i}z{\i}lo{\v g}lu}, {\"U}., {K{\i}z{\i}lo{\v g}lu}, N.,
  {Balman}, {\c S}., \& {Inam}, S.~{\c C}. 2005, \aap, 439, 1131

\bibitem[{{Bell} {et~al.}(1993){Bell}, {Hilditch}, \& {Pollacco}}]{bell93}
{Bell}, S.~A., {Hilditch}, R.~W., \& {Pollacco}, D.~L. 1993, \mnras, 265, 1042

\bibitem[{{Bessell}(1990)}]{bessel90}
{Bessell}, M.~S. 1990, \pasp, 102, 1181

\bibitem[{{Bonnarel} {et~al.}(2000){Bonnarel}, {Fernique}, {Bienaym{\'e}},
  {Egret}, {Genova}, {Louys}, {Ochsenbein}, {Wenger}, \&
  {Bartlett}}]{bonnarel00}
{Bonnarel}, F., {Fernique}, P., {Bienaym{\'e}}, O., {et~al.} 2000, \aaps, 143,
  33

\bibitem[{{Bosch-Ramon}(2013)}]{bosch13}
{Bosch-Ramon}, V. 2013, \aap, 560, A32

\bibitem[{{Dachs} {et~al.}(1988){Dachs}, {Kiehling}, \& {Engels}}]{dachs88}
{Dachs}, J., {Kiehling}, R., \& {Engels}, D. 1988, \aap, 194, 167

\bibitem[{{Fabregat} \& {Reglero}(1990)}]{fabregat90}
{Fabregat}, J. \& {Reglero}, V. 1990, \mnras, 247, 407

\bibitem[{{Finley} {et~al.}(1994){Finley}, {Taylor}, \& {Belloni}}]{finley94}
{Finley}, J.~P., {Taylor}, M., \& {Belloni}, T. 1994, \apj, 429, 356

\bibitem[{{Fitzgerald}(1970)}]{fitzgerald70}
{Fitzgerald}, M.~P. 1970, \aap, 4, 234

\bibitem[{{Gehrz} {et~al.}(1974){Gehrz}, {Hackwell}, \& {Jones}}]{gehrz74}
{Gehrz}, R.~D., {Hackwell}, J.~A., \& {Jones}, T.~W. 1974, \apj, 191, 675

\bibitem[{{Goranskii}(2001)}]{goranskii01}
{Goranskii}, V.~P. 2001, Astronomy Letters, 27, 516

\bibitem[{{Grimm} {et~al.}(2003){Grimm}, {Gilfanov}, \& {Sunyaev}}]{grimm03}
{Grimm}, H.-J., {Gilfanov}, M., \& {Sunyaev}, R. 2003, \mnras, 339, 793

\bibitem[{{Gutierrez-Moreno}(1979)}]{gutierrez-moreno79}
{Gutierrez-Moreno}, A. 1979, \pasp, 91, 299

\bibitem[{{Guti{\'e}rrez-Soto} {et~al.}(2011){Guti{\'e}rrez-Soto}, {Reig},
  {Fabregat}, \& {Fox-Machado}}]{gutierrez-soto11}
{Guti{\'e}rrez-Soto}, J., {Reig}, P., {Fabregat}, J., \& {Fox-Machado}, L.
  2011, in IAU Symposium, Vol. 272, IAU Symposium, ed. C.~{Neiner}, G.~{Wade},
  G.~{Meynet}, \& G.~{Peters}, 505--506

\bibitem[{{Hadrava} \& {{\v C}echura}(2012)}]{hadrava12}
{Hadrava}, P. \& {{\v C}echura}, J. 2012, \aap, 542, A42

\bibitem[{{Haigh} {et~al.}(1999){Haigh}, {Coe}, {Steele}, \&
  {Fabregat}}]{haigh99}
{Haigh}, N.~J., {Coe}, M.~J., {Steele}, I.~A., \& {Fabregat}, J. 1999, \mnras,
  310, L21

\bibitem[{{Harmanec}(1983)}]{harmanec83}
{Harmanec}, P. 1983, Hvar Observatory Bulletin, 7, 55

\bibitem[{{Harmanec}(2000)}]{harmanec00}
{Harmanec}, P. 2000, in Astronomical Society of the Pacific Conference Series,
  Vol. 214, IAU Colloq. 175: The Be Phenomenon in Early-Type Stars, ed. M.~A.
  {Smith}, H.~F. {Henrichs}, \& J.~{Fabregat}, 13

\bibitem[{{Haubois} {et~al.}(2012){Haubois}, {Carciofi}, {Rivinius}, {Okazaki},
  \& {Bjorkman}}]{haubois12}
{Haubois}, X., {Carciofi}, A.~C., {Rivinius}, T., {Okazaki}, A.~T., \&
  {Bjorkman}, J.~E. 2012, \apj, 756, 156

\bibitem[{{Jaschek} \& {G{\'o}mez}(1998)}]{jaschek98}
{Jaschek}, C. \& {G{\'o}mez}, A.~E. 1998, Highlights of Astronomy, 11, 566

\bibitem[{{Johnson}(1966)}]{johnson66}
{Johnson}, H.~L. 1966, \araa, 4, 193

\bibitem[{{Jones} {et~al.}(2008){Jones}, {Sigut}, \& {Porter}}]{jones08}
{Jones}, C.~E., {Sigut}, T.~A.~A., \& {Porter}, J.~M. 2008, \mnras, 386, 1922

\bibitem[{{K{\i}z{\i}lo{\v g}lu} {et~al.}(2007){K{\i}z{\i}lo{\v g}lu},
  {K{\i}z{\i}lo{\v g}lu}, {Baykal}, {Yerli}, \& {{\"O}zbey}}]{kiziloglu07b}
{K{\i}z{\i}lo{\v g}lu}, U., {K{\i}z{\i}lo{\v g}lu}, N., {Baykal}, A., {Yerli},
  S.~K., \& {{\"O}zbey}, M. 2007, \aap, 470, 1023

\bibitem[{{K{\i}z{\i}lo{\v g}lu} {et~al.}(2009){K{\i}z{\i}lo{\v g}lu},
  {{\"O}zbilgen}, {K{\i}z{\i}lo{\v g}lu}, \& {Baykal}}]{kiziloglu09}
{K{\i}z{\i}lo{\v g}lu}, {\"U}., {{\"O}zbilgen}, S., {K{\i}z{\i}lo{\v g}lu}, N.,
  \& {Baykal}, A. 2009, \aap, 508, 895

\bibitem[{{Kretschmar} {et~al.}(2012){Kretschmar}, {Nespoli}, {Reig}, \&
  {Anders}}]{kretschmar12}
{Kretschmar}, P., {Nespoli}, E., {Reig}, P., \& {Anders}, F. 2012, in
  Proceedings of ''An INTEGRAL view of the high-energy sky (the first 10
  years)'' - 9th INTEGRAL Workshop and celebration of the 10th anniversary of
  the launch (INTEGRAL 2012). 15-19 October 2012. Bibliotheque Nationale de
  France, Paris, France. Published online at <A
  href=''http://pos.sissa.it/cgi-bin/reader/conf.cgi?confid=176''>http://pos.s%
issa.it/cgi-bin/reader/conf.cgi?confid=176</A>, id.16

\bibitem[{{Landolt}(1992)}]{landolt92}
{Landolt}, A.~U. 1992, \aj, 104, 340

\bibitem[{{Landolt}(2009)}]{landolt09}
{Landolt}, A.~U. 2009, \aj, 137, 4186

\bibitem[{{Larionov} {et~al.}(2001){Larionov}, {Lyuty}, \&
  {Zaitseva}}]{larionov01}
{Larionov}, V., {Lyuty}, V.~M., \& {Zaitseva}, G.~V. 2001, \aap, 378, 837

\bibitem[{{Lyuty} \& {Za{\u i}tseva}(2000)}]{lyuty00}
{Lyuty}, V.~M. \& {Za{\u i}tseva}, G.~V. 2000, Astronomy Letters, 26, 9

\bibitem[{{Manousakis} {et~al.}(2012){Manousakis}, {Walter}, \&
  {Blondin}}]{manousakis12}
{Manousakis}, A., {Walter}, R., \& {Blondin}, J.~M. 2012, \aap, 547, A20

\bibitem[{{Mendelson} \& {Mazeh}(1991)}]{mendelson91}
{Mendelson}, H. \& {Mazeh}, T. 1991, \mnras, 250, 373

\bibitem[{{Mineo} {et~al.}(2012){Mineo}, {Gilfanov}, \& {Sunyaev}}]{mineo12}
{Mineo}, S., {Gilfanov}, M., \& {Sunyaev}, R. 2012, \mnras, 419, 2095

\bibitem[{{Negueruela}(2010)}]{negueruela10}
{Negueruela}, I. 2010, in Astronomical Society of the Pacific Conference
  Series, Vol. 422, High Energy Phenomena in Massive Stars, ed.
  J.~{Mart{\'{\i}}}, P.~L. {Luque-Escamilla}, \& J.~A. {Combi}, 57

\bibitem[{{Negueruela} \& {Okazaki}(2001)}]{negueruela01}
{Negueruela}, I. \& {Okazaki}, A.~T. 2001, \aap, 369, 108

\bibitem[{{Reig} {et~al.}(2007){Reig}, {Larionov}, {Negueruela}, {Arkharov}, \&
  {Kudryavtseva}}]{reig07b}
{Reig}, P., {Larionov}, V., {Negueruela}, I., {Arkharov}, A.~A., \&
  {Kudryavtseva}, N.~A. 2007, \aap, 462, 1081

\bibitem[{{Reig} \& {Zezas}(2014)}]{reig14a}
{Reig}, P. \& {Zezas}, A. 2014, \aap, 561, A137

\bibitem[{{Rivinius} {et~al.}(2013){Rivinius}, {Carciofi}, \&
  {Martayan}}]{rivinius13a}
{Rivinius}, T., {Carciofi}, A.~C., \& {Martayan}, C. 2013, \aapr, 21, 69

\bibitem[{{Sarty} {et~al.}(2009){Sarty}, {Kiss}, {Huziak}, {Catalan}, {Luciuk},
  {Crawford}, {Lane}, {Pickard}, {Grzybowski}, {Closas}, {Johnston}, {Balam},
  \& {Wu}}]{sarty09}
{Sarty}, G.~E., {Kiss}, L.~L., {Huziak}, R., {et~al.} 2009, \mnras, 392, 1242

\bibitem[{{Sigut} \& {Patel}(2013)}]{sigut13}
{Sigut}, T.~A.~A. \& {Patel}, P. 2013, \apj, 765, 41

\bibitem[{{Telting} {et~al.}(1998){Telting}, {Waters}, {Roche}, {Boogert},
  {Clark}, {de Martino}, \& {Persi}}]{telting98}
{Telting}, J.~H., {Waters}, L.~B.~F.~M., {Roche}, P., {et~al.} 1998, \mnras,
  296, 785

\bibitem[{{Tomsick} \& {Muterspaugh}(2010)}]{tomsick10}
{Tomsick}, J.~A. \& {Muterspaugh}, M.~W. 2010, \apj, 719, 958

\bibitem[{{Torrej{\'o}n} {et~al.}(2007){Torrej{\'o}n}, {Negueruela}, \&
  {Riquelme}}]{torrejon07}
{Torrej{\'o}n}, J.~M., {Negueruela}, I., \& {Riquelme}, M.~S. 2007, in
  Astronomical Society of the Pacific Conference Series, Vol. 361, Active
  OB-Stars: Laboratories for Stellare and Circumstellar Physics, ed. A.~T.
  {Okazaki}, S.~P. {Owocki}, \& S.~{Stefl}, 503

\bibitem[{{van der Meer} {et~al.}(2007){van der Meer}, {Kaper}, {van Kerkwijk},
  {Heemskerk}, \& {van den Heuvel}}]{meer07}
{van der Meer}, A., {Kaper}, L., {van Kerkwijk}, M.~H., {Heemskerk}, M.~H.~M.,
  \& {van den Heuvel}, E.~P.~J. 2007, \aap, 473, 523

\bibitem[{{Waters} {et~al.}(1988){Waters}, {van den Heuvel}, {Taylor},
  {Habets}, \& {Persi}}]{waters88}
{Waters}, L.~B.~F.~M., {van den Heuvel}, E.~P.~J., {Taylor}, A.~R., {Habets},
  G.~M.~H.~J., \& {Persi}, P. 1988, \aap, 198, 200

\bibitem[{{Wegner}(1994)}]{wegner94}
{Wegner}, W. 1994, \mnras, 270, 229

\bibitem[{{Wegner}(2006)}]{wegner06}
{Wegner}, W. 2006, \mnras, 371, 185

\end{thebibliography}

\begin{appendix}

\section{Secondary standard stars}

Table~\ref{secstd1} to ~\ref{secstd5} list the final set of secondary
standard stars for each target. In these tables, column 5 gives the angular
distance between the star and the target. Columns 6--9 are the mean
magnitudes, columns 10--13 show the standard deviation of the mean
calculated from eq.~(\ref{emean}) and columns 14--17 show the number of
measurements considered. The RA and DEC and the angular distance from the
target to the secondary standards were derived using the ALADIN Sky Atlas
\citep{bonnarel00}. 

Finding charts with the identification of the secondary
standard stars are also available in electronic form at the CDS through the
VizieR service http://vizier.u-strasbg.fr/viz-bin/VizieR.


\begin{sidewaystable*}
\caption{Secondary standards in the field of view of high-mass X-ray
binaries. The mean and standard deviation of the mean in the
$BVRI$ bands (calculated from eqs. (\ref{mean}) and (\ref{emean})) and the
number of observations are given.}
\label{secstd1}
\begin{center}
\begin{tabular}{lcccccccccccccccc}
\hline  \hline
Star &ID &RA    &DEC            &d      &$\overline{B}$ &$\overline{V}$ &$\overline{R}$ &$\overline{I}$ &$\sigma_{\overline B}$     &$\sigma_{\overline V}$  &$\sigma_{\overline R}$ &$\sigma_{\overline I}$ &$N_B$      &$N_V$  &$N_R$  &$N_I$  \\
     &   &      &               &(\arcmin)      &(mag)  &(mag)  &(mag)  &(mag)  &(mag)          &(mag)      &(mag)       &(mag)                &        &        &       &       \\     
\hline\hline
\multicolumn{16}{c}{2S\,0114+65} \\
\hline
C1  &9    &01 18 19.3 &+65 20 28  &3.4   &14.908&13.903&13.274&12.629&0.017&0.004&0.008&0.016&9&9&9&7    \\
C2  &26   &01 17 29.4 &+65 17 06  &3.5   &11.933&11.256&10.877&10.419&0.006&0.009&0.004&0.006&10&10&10&8 \\
C3  &36   &01 17 31.9 &+65 15 59  &3.6   &15.370&13.991&13.095&12.318&0.013&0.011&0.008&0.008&10&10&10&8 \\
C4  &47   &01 17 54.7 &+65 14 36  &3.0   &13.846&13.029&12.546&12.091&0.008&0.005&0.005&0.006&10&10&10&8 \\
\hline  
\multicolumn{16}{c}{4U\,0115+63} \\
\hline
C1  &7   &01 18 45.6 &+63 47 57  &3.7  &16.606&15.397&14.682&14.040&0.011&0.008&0.007&0.006&22&22&22&17 \\
C2  &8   &01 18 19.2 &+63 47 56  &3.6  &14.629&13.755&13.255&12.708&0.008&0.008&0.007&0.006&22&22&22&18 \\
C3  &9   &01 18 58.9 &+63 47 53  &4.4  &17.095&15.829&15.092&14.337&0.009&0.008&0.008&0.008&22&23&22&18 \\
C4  &10  &01 18 46.9 &+63 47 49  &3.6  &16.734&15.358&14.541&13.605&0.007&0.008&0.008&0.008&22&23&23&18 \\
C5  &12  &01 18 16.6 &+63 47 38  &3.5  &16.583&15.440&14.768&14.089&0.008&0.007&0.008&0.007&22&23&23&19 \\
C6  &15  &01 18 31.4 &+63 47 29  &2.9  &16.332&15.066&14.337&13.470&0.008&0.008&0.007&0.007&22&23&24&18 \\
C7  &32  &01 18 46.4 &+63 46 03  &2.1  &16.121&14.612&13.711&12.699&0.008&0.008&0.007&0.008&23&25&24&21 \\
C8  &54  &01 18 57.7 &+63 44 59  &2.8  &14.347&13.535&13.108&12.561&0.007&0.008&0.007&0.004&22&23&23&17 \\
C9  &62  &01 18 49.2 &+63 44 22  &2.0  &15.464&14.440&13.853&13.195&0.007&0.008&0.007&0.005&24&25&25&20 \\
C10 &67  &01 18 45.1 &+63 44 04  &1.5  &13.737&13.005&12.619&12.127&0.011&0.009&0.009&0.007&18&17&18&13 \\
C11 &98  &01 18 10.8 &+63 41 44  &3.6  &17.142&15.934&15.231&14.398&0.008&0.008&0.008&0.008&22&25&25&21 \\
\hline  
\multicolumn{16}{c}{IGR\,J01363+6610} \\
\hline
C1  &12  &01 36 20.6 &66 15 34  &4.1  &17.101&15.281&14.221&13.192&0.014&0.008&0.005&0.009&11&11&11&11 \\
C2  &13  &01 36 09.6 &66 15 17  &3.2  &16.060&14.806&14.071&13.338&0.011&0.004&0.004&0.005&12&11&12&12 \\
C3  &17  &01 36 20.5 &66 14 41  &3.6  &16.271&15.017&14.265&13.457&0.010&0.008&0.006&0.004&12&13&13&12 \\
C4  &18  &01 36 28.1 &66 14 31  &4.2  &15.246&13.375&12.287&11.255&0.013&0.011&0.008&0.007&12&13&13&12 \\
C5  &75  &01 35 53.2 &66 11 27  &1.2  &16.284&14.877&14.068&13.311&0.009&0.006&0.004&0.006&16&15&15&15 \\
C6  &82  &01 36 20.4 &66 10 14  &3.9  &15.743&14.582&13.895&13.196&0.010&0.007&0.006&0.008&14&14&14&14 \\
C7  &87  &01 35 34.6 &66 09 25  &3.6  &16.808&15.340&14.366&13.422&0.012&0.005&0.005&0.008&16&15&15&15 \\
C8  &90  &01 35 18.5 &66 08 38  &5.1  &14.361&13.191&12.541&11.920&0.011&0.005&0.005&0.004&13&12&12&12 \\
\hline  
\multicolumn{16}{c}{IGR\,J01583+6713} \\
\hline
C1  &23  &01 58 22.3 &+67 14 33  &1.2  &17.743&16.369&15.553&14.814&0.014&0.008&0.008&0.009&9&9&9&9 \\
C2  &24  &01 58 20.0 &+67 14 29  &1.1  &16.192&14.886&14.094&13.315&0.014&0.005&0.006&0.006&9&9&9&9 \\
C3  &28  &01 57 55.7 &+67 13 58  &2.2  &16.299&14.871&14.019&13.119&0.014&0.005&0.006&0.005&9&9&9&9 \\
C4  &29  &01 57 45.1 &+67 13 48  &3.2  &17.546&16.240&15.473&14.605&0.015&0.004&0.006&0.004&9&9&9&9 \\
C5  &55  &01 58 30.9 &+67 13 20  &1.2  &16.039&14.682&13.863&13.144&0.015&0.013&0.013&0.013&9&9&9&9 \\
C6  &85  &01 58 30.9 &+67 10 35  &3.0  &15.361&13.992&13.201&12.339&0.010&0.005&0.005&0.005&9&9&9&9 \\
C7  &90  &01 57 47.4 &+67 10 06  &4.4  &16.960&15.589&14.658&13.863&0.014&0.005&0.006&0.006&9&9&9&9 \\
C8  &93  &01 58 24.8 &+67 10 03  &3.3  &17.782&16.254&15.206&13.780&0.015&0.006&0.006&0.005&9&9&9&9 \\
\hline\hline
\end{tabular}
\end{center}
\end{sidewaystable*}

\begin{sidewaystable*}
\caption{Secondary standards in the field of view of high-mass X-ray
binaries (cont.)}
\label{secstd2}
\begin{center}
\begin{tabular}{lcccccccccccccccc}
\hline  \hline
Star &ID &RA    &DEC            &d      &$\overline{B}$ &$\overline{V}$ &$\overline{R}$ &$\overline{I}$ &$\sigma_{\overline B}$     &$\sigma_{\overline V}$  &$\sigma_{\overline R}$ &$\sigma_{\overline I}$ &$N_B$      &$N_V$  &$N_R$  &$N_I$  \\
     &   &      &               &(\arcmin)      &(mag)  &(mag)  &(mag)  &(mag)  &(mag)          &(mag)      &(mag)       &(mag)                &        &        &       &       \\     
\hline  \hline
\multicolumn{16}{c}{RX\,J0146.9+6121} \\
\hline
C1  &9   &01 47 30.0 &+61 25 18  &5.3  &12.623&12.041&11.676&11.315&0.007&0.004&0.005&0.007&18&17&17&15 \\
C2  &12  &01 47 39.4 &+61 24 43  &4.2  &14.033&13.061&12.506&11.975&0.006&0.007&0.006&0.005&20&23&21&18 \\
C3  &13  &01 47 19.6 &+61 24 23  &3.8  &14.174&13.381&12.938&12.434&0.006&0.006&0.005&0.005&22&22&22&17 \\
C4  &14  &01 47 30.4 &+61 23 14  &4.0  &13.250&12.608&12.226&11.780&0.007&0.006&0.006&0.005&20&20&20&17 \\
C5  &16  &01 47 22.5 &+61 23 07  &3.2  &14.713&13.710&13.131&12.483&0.007&0.007&0.005&0.006&22&22&22&19 \\
C6  &24  &01 47 15.2 &+61 22 12  &1.9  &13.525&12.900&12.525&12.092&0.006&0.006&0.005&0.003&24&24&24&18 \\
C7  &34  &01 46 32.6 &+61 21 18  &3.3  &12.496&11.681&11.211&10.652&0.006&0.005&0.006&0.004&23&23&23&20 \\
C8  &35  &01 47 22.8 &+61 21 02  &2.6  &11.388&10.841&10.512&10.123&0.008&0.005&0.005&0.004&22&20&21&17 \\
C9  &39  &01 46 51.6 &+61 20 54  &1.1  &14.532&13.736&13.255&12.750&0.007&0.005&0.006&0.004&25&25&24&20 \\
C10 &48  &01 46 47.7 &+61 20 09  &1.8  &12.295&11.501&11.042&10.509&0.007&0.004&0.005&0.004&25&23&24&19 \\
C11 &53  &01 46 51.4 &+61 19 28  &2.1  &13.028&12.348&11.944&11.543&0.006&0.003&0.005&0.005&24&22&24&20 \\
C12 &55  &01 47 20.9 &+61 19 10  &3.2  &14.057&13.377&13.001&12.529&0.008&0.006&0.006&0.005&26&23&24&20 \\
C13 &64  &01 47 09.5 &+61 18 42  &2.8  &13.577&12.568&11.995&11.444&0.008&0.005&0.005&0.005&25&24&23&20 \\
C14 &78  &01 47 23.8 &+61 17 45  &4.5  &14.904&13.912&13.368&12.734&0.011&0.006&0.009&0.012&22&19&21&18 \\
C15 &85  &01 46 41.9 &+61 27 14  &4.6  &13.591&12.981&12.619&12.188&0.010&0.009&0.010&0.004&14&14&14&12 \\
\hline  
\multicolumn{16}{c}{RX\,J0240.4+6112} \\
\hline
C1  &35  &02 40 43.6 &+61 16 21  &2.8  &13.450&12.875&12.547&12.141&0.012&0.012&0.010&0.024&8&8&8&8      \\
C2  &39  &02 40 51.0 &+61 15 45  &2.9  &12.177&11.615&11.274&10.888&0.005&0.009&0.007&0.014&9&10&10&9    \\
C3  &41  &02 40 35.1 &+61 15 37  &1.8  &14.016&13.345&12.947&12.475&0.009&0.010&0.011&0.013&10&11&10&10  \\
C4  &45  &02 39 59.1 &+61 15 19  &4.2  &11.151&10.798&10.604&10.341&0.011&0.009&0.009&0.012&10&10&9&9    \\
C5  &47  &02 40 24.1 &+61 15 14  &1.7  &15.032&14.244&13.812&13.289&0.012&0.010&0.008&0.006&11&11&10&10  \\
C6  &59  &02 40 13.1 &+61 13 39  &2.1  &13.821&12.869&12.324&11.746&0.008&0.011&0.011&0.014&10&11&10&10  \\
C7  &64  &02 40 35.7 &+61 12 43  &1.1  &13.171&12.392&11.957&11.426&0.007&0.009&0.006&0.007&10&11&10&10  \\
C8  &72  &02 40 55.8 &+61 11 20  &3.7  &14.103&13.322&12.873&12.401&0.009&0.012&0.009&0.012&9&9&9&8      \\
C9  &80  &02 39 57.4 &+61 10 48  &5.0  &16.129&15.146&14.600&13.989&0.012&0.008&0.009&0.012&9&10&9&10    \\
C10 &81  &02 40 10.7 &+61 10 45  &3.8  &14.571&13.901&13.514&13.077&0.006&0.009&0.011&0.016&10&11&10&10  \\
C11 &84  &02 39 57.1 &+61 10 28  &5.2  &15.088&14.180&13.637&13.079&0.016&0.010&0.011&0.009&10&10&9&10   \\
C12 &86  &02 40 49.5 &+61 10 25  &3.9  &14.678&13.393&12.675&11.970&0.008&0.012&0.012&0.010&10&11&10&10  \\
\hline                                                                                                   
\multicolumn{16}{c}{V\,0332+53} \\
\hline
C1  &27  &03 34 37.2 &+53 12 28  &3.9  &17.966&16.395&15.472&14.538&0.013&0.012&0.013&0.007&9&9&9&7 \\
C2  &32  &03 35 03.7 &+53 12 09  &1.7  &14.908&13.790&13.128&12.477&0.011&0.012&0.010&0.004&9&9&9&7 \\
C3  &34  &03 34 52.9 &+53 11 53  &1.7  &16.425&14.897&13.979&13.045&0.009&0.010&0.010&0.006&9&9&9&7 \\
C4  &45  &03 34 36.6 &+53 10 23  &3.4  &17.035&15.632&14.813&13.973&0.015&0.011&0.012&0.007&9&9&9&7 \\
C5  &48  &03 34 50.1 &+53 09 58  &1.5  &16.924&15.493&14.522&13.417&0.012&0.014&0.013&0.008&9&9&9&7 \\
C6  &57  &03 34 40.9 &+53 08 36  &3.3  &16.861&15.240&14.225&13.140&0.013&0.010&0.009&0.008&9&9&9&7 \\
\hline\hline
\end{tabular}
\end{center}
\end{sidewaystable*}

\begin{sidewaystable*}
\caption{Secondary standards in the field of view of high-mass X-ray
binaries (cont.)}
\label{secstd3}
\begin{center}
\begin{tabular}{lcccccccccccccccc}
\hline  \hline
Star &ID &RA    &DEC            &d      &$\overline{B}$ &$\overline{V}$ &$\overline{R}$ &$\overline{I}$ &$\sigma_{\overline B}$     &$\sigma_{\overline V}$  &$\sigma_{\overline R}$ &$\sigma_{\overline I}$ &$N_B$      &$N_V$  &$N_R$  &$N_I$  \\
     &   &      &               &(\arcmin)&(mag)        &(mag)          &(mag)         &(mag)          &(mag)                  &(mag)                  &(mag)                 &(mag)          &        &       &       &       \\     

\hline  \hline
\multicolumn{16}{c}{RX\,J0440.9+4431} \\
\hline
C1  &3   &04 41 05.0 &+44 34 39  &3.0  &15.503&14.437&13.740&12.905&0.012&0.013&0.012&0.004&13&13&13&11 \\
C2  &4   &04 40 45.1 &+44 34 12  &3.4  &13.410&12.334&11.710&11.093&0.007&0.009&0.007&0.004&13&13&13&12 \\
C3  &6   &04 41 25.0 &+44 33 25  &4.8  &15.550&14.426&13.751&13.040&0.016&0.014&0.016&0.008&8&8&8&8     \\
C4  &7   &04 40 54.9 &+44 33 18  &1.6  &14.879&13.926&13.367&12.757&0.007&0.012&0.011&0.009&15&15&15&14 \\
C5  &9   &04 40 43.4 &+44 32 29  &2.8  &15.665&14.625&14.022&13.403&0.012&0.010&0.010&0.009&14&14&14&13 \\
C6  &12  &04 40 35.6 &+44 32 04  &4.2  &15.688&14.738&14.181&13.587&0.010&0.015&0.004&0.009&10&10&10&9  \\
C7  &13  &04 40 37.3 &+44 31 58  &3.8  &14.524&13.758&13.315&12.800&0.010&0.012&0.009&0.010&12&12&12&11 \\
C8  &18  &04 40 40.9 &+44 31 24  &3.2  &14.707&13.696&13.089&12.437&0.006&0.008&0.008&0.006&13&13&13&12 \\
C9  &19  &04 40 35.9 &+44 31 18  &4.0  &15.454&14.504&13.927&13.323&0.008&0.012&0.009&0.005&12&12&12&10 \\
C10 &20  &04 40 39.2 &+44 31 04  &3.6  &15.560&14.818&14.400&13.884&0.013&0.012&0.013&0.014&13&13&13&12 \\
C11 &23  &04 40 59.4 &+44 30 07  &1.6  &14.444&13.717&13.288&12.774&0.010&0.009&0.013&0.017&15&15&15&14 \\
\hline  
\multicolumn{16}{c}{1A\,0535+262} \\
\hline
C1  &2   &05 39 09.6 &+26 22 48  &5.0  &11.047&10.754&10.578&10.327&0.013&0.009&0.014&0.019&7&7&7&6    \\
C2  &4   &05 39 09.5 &+26 22 25  &4.7  &10.211&10.081&9.994&9.878&0.014&0.010&0.014&0.018&7&7&7&6      \\
C3  &7   &05 38 56.9 &+26 21 27  &2.5  &14.029&13.240&12.788&12.327&0.012&0.012&0.014&0.011&7&7&7&6    \\
C4  &15  &05 39 01.9 &+26 20 34  &2.2  &14.735&13.802&13.256&12.656&0.012&0.009&0.016&0.022&7&7&7&6    \\
C5  &17  &05 38 47.6 &+26 20 17  &2.0  &13.231&12.705&12.389&12.001&0.010&0.009&0.011&0.014&7&7&7&6    \\
C6  &26  &05 39 09.2 &+26 18 42  &3.2  &14.080&13.355&12.940&12.528&0.014&0.011&0.013&0.019&7&7&7&6     \\
C7  &33  &05 38 41.6 &+26 17 17  &3.2  &14.761&13.981&13.544&13.101&0.016&0.012&0.016&0.023&7&7&7&6     \\
\hline  
\multicolumn{16}{c}{IGR\,J06074+2205} \\
\hline
C1  &18  &06 07 12.0 &+22 07 42  &3.9  &15.033&14.450&14.145&13.733&0.019&0.011&0.017&0.019&6&6&6&6     \\
C2  &22  &06 07 14.8 &+22 07 18  &3.0  &14.711&14.011&13.615&13.192&0.021&0.017&0.016&0.026&6&6&6&6     \\
C3  &25  &06 07 32.2 &+22 07 04  &1.8  &14.522&13.717&13.275&12.859&0.017&0.014&0.013&0.022&6&6&6&6     \\
C4  &51  &06 07 22.8 &+22 05 03  &1.1  &14.856&14.203&13.817&13.417&0.021&0.015&0.015&0.028&6&6&6&6     \\
C5  &64  &06 07 15.3 &+22 03 35  &3.4  &14.695&14.077&13.750&13.325&0.023&0.016&0.018&0.020&6&6&6&6     \\
C6  &66  &06 07 26.3 &+22 03 06  &2.6  &14.611&13.914&13.513&13.102&0.013&0.015&0.014&0.018&6&6&6&6     \\
C7  &75  &06 07 20.6 &+22 01 53  &4.2  &14.625&14.151&13.878&13.526&0.021&0.007&0.014&0.024&6&6&6&6     \\
C8  &79  &06 07 36.6 &+22 01 15  &5.0  &15.185&14.411&13.970&13.512&0.020&0.020&0.013&0.025&6&6&6&6     \\
\hline  
\multicolumn{16}{c}{4U\,J1845.0-0433} \\
\hline
C1  &13  &18 45 04.5 &--04 30 32  &3.4  &16.153&15.082&14.439&13.768&0.012&0.010&0.007&0.011&10&10&10&9    \\
C2  &19  &18 45 02.9 &--04 31 27  &2.5  &15.394&14.298&13.661&12.940&0.014&0.013&0.007&0.005&11&11&10&9    \\
C3  &27  &18 45 06.6 &--04 31 50  &2.4  &16.476&14.651&13.518&12.402&0.014&0.009&0.009&0.009&11&11&11&9    \\
C4  &33  &18 44 49.0 &--04 32 19  &3.5  &16.183&15.008&14.300&13.498&0.013&0.008&0.007&0.007&11&11&11&9    \\
C5  &39  &18 44 49.5 &--04 32 45  &3.1  &15.107&13.958&13.266&12.472&0.013&0.008&0.007&0.006&11&11&11&9    \\
C6  &42  &18 45 01.1 &--04 33 07  &0.8  &15.647&14.620&14.013&13.336&0.011&0.010&0.007&0.011&11&11&11&10   \\
C7  &47  &18 45 10.3 &--04 33 38  &2.1  &15.226&14.151&13.496&12.803&0.011&0.010&0.007&0.011&11&11&11&10   \\
C8  &55  &18 45 08.3 &--04 34 13  &1.6  &16.651&15.126&14.115&13.058&0.012&0.008&0.008&0.008&11&11&11&9    \\
C9  &56  &18 45 16.1 &--04 34 19  &3.6  &15.914&14.841&14.188&13.508&0.012&0.008&0.008&0.011&10&10&10&10   \\
C10 &59  &18 45 12.1 &--04 34 22  &2.6  &16.707&14.813&13.728&12.698&0.013&0.009&0.009&0.009&11&11&11&9    \\
C11 &79  &18 44 48.7 &--04 36 23  &3.9  &16.292&15.168&14.487&13.753&0.014&0.009&0.008&0.018&10&10&10&10   \\
C12 &89  &18 44 56.8 &--04 37 06  &3.2  &17.287&15.162&13.827&12.559&0.014&0.008&0.009&0.009&10&10&11&9    \\
\hline\hline
\end{tabular}
\end{center}
\end{sidewaystable*}

\begin{sidewaystable*}
\caption{Secondary standards in the field of view of high-mass X-ray
binaries (cont.)}
\label{secstd4}
\begin{center}
\begin{tabular}{lcccccccccccccccc}
\hline  \hline

Star &ID &RA    &DEC            &d      &$\overline{B}$ &$\overline{V}$ &$\overline{R}$ &$\overline{I}$ &$\sigma_{\overline B}$     &$\sigma_{\overline V}$  &$\sigma_{\overline R}$ &$\sigma_{\overline I}$ &$N_B$      &$N_V$  &$N_R$  &$N_I$  \\
     &   &      &               &(\arcmin)&(mag)        &(mag)          &(mag)         &(mag)          &(mag)                  &(mag)                  &(mag)                 &(mag)          &        &       &       &       \\     

\hline  \hline                                                  
\multicolumn{16}{c}{4U\,1907+09} \\
\hline
C1  &41  &19 09 43.2 &+09 51 59  &2.5  &18.522&16.935&15.971&15.100&0.008&0.005&0.008&0.007&11&11&12&8   \\
C2  &42  &19 09 40.2 &+09 51 51  &2.1  &18.873&17.103&16.050&14.920&0.012&0.010&0.013&0.015&12&11&12&8   \\
C3  &47  &19 09 27.4 &+09 51 06  &2.8  &17.018&15.383&14.461&13.533&0.009&0.005&0.007&0.007&13&13&13&10  \\
C4  &52  &19 09 26.7 &+09 50 53  &2.9  &17.795&16.203&15.274&14.257&0.013&0.011&0.008&0.013&12&13&13&10  \\
C5  &55  &19 09 31.7 &+09 50 15  &1.6  &18.191&16.759&15.908&15.014&0.010&0.011&0.009&0.010&12&13&13&10  \\
C6  &70  &19 09 37.6 &+09 49 54  &0.1  &17.608&16.034&15.137&14.208&0.011&0.006&0.008&0.007&13&13&14&10  \\
C7  &77  &19 09 38.0 &+09 49 47  &1.6  &17.214&15.281&14.166&13.095&0.009&0.010&0.007&0.010&12&13&13&10  \\
C8  &119 &19 09 23.0 &+09 47 25  &4.3  &18.419&16.995&16.126&15.314&0.013&0.007&0.010&0.012&9&9&9&7      \\
C9  &142 &19 09 28.7 &+09 47 03  &3.5  &18.104&16.545&15.501&14.560&0.015&0.004&0.011&0.009&10&10&10&8  \\
C10 &146 &19 09 37.0 &+09 46 40  &3.1  &17.196&15.541&14.559&13.477&0.010&0.004&0.008&0.007&11&11&11&8  \\
\hline  \hline                                                  
\multicolumn{16}{c}{XTE\,J1946+274} \\
\hline
C1  &55   &19 45 44.7 &+27 24 27  &2.7  &14.969&14.109&13.607&13.117&0.010&0.006&0.006&0.009&7&7&7&7 \\
C2  &62   &19 45 54.2 &+27 24 13  &4.0  &15.723&14.309&13.506&12.746&0.011&0.006&0.007&0.009&7&7&7&7 \\
C3  &92   &19 45 56.8 &+27 22 50  &3.9  &17.499&15.666&14.590&13.587&0.011&0.014&0.008&0.011&7&7&7&7 \\
C4  &97   &19 45 58.3 &+27 22 38  &4.2  &16.411&15.443&14.862&14.294&0.008&0.005&0.010&0.007&7&7&7&7 \\
C5  &108  &19 45 44.4 &+27 22 18  &1.1  &16.554&15.053&14.217&13.456&0.010&0.007&0.009&0.013&7&7&7&7 \\
C6  &137  &19 45 27.7 &+27 20 52  &2.7  &16.304&15.443&14.911&14.363&0.009&0.007&0.010&0.006&7&7&7&7 \\
C7  &140  &19 45 38.5 &+27 20 47  &1.1  &17.156&15.530&14.607&13.732&0.015&0.011&0.013&0.015&7&7&7&7 \\
C8  &161  &19 45 43.0 &+27 19 44  &2.3  &15.235&14.354&13.822&13.291&0.008&0.005&0.007&0.010&7&7&7&7 \\
C9  &164  &19 45 37.4 &+27 19 40  &2.3  &17.185&15.144&13.948&12.823&0.012&0.010&0.011&0.023&7&7&7&7 \\
C10 &166  &19 45 45.3 &+27 19 38  &2.6  &16.886&15.243&14.275&13.371&0.018&0.008&0.009&0.011&7&7&7&7 \\
C11 &167  &19 45 46.9 &+27 19 35  &2.8  &16.984&15.881&15.271&14.670&0.012&0.012&0.008&0.015&7&7&7&7 \\
\hline  \hline                                                  
\multicolumn{16}{c}{KS\,1947+300} \\
\hline
C1  &99   &19 49 41.5 &+30 13 47  &1.8  &15.329&14.504&14.002&13.537&0.006&0.006&0.006&0.009&17&17&17&14 \\
C2  &103  &19 49 18.4 &+30 13 42  &3.8  &15.129&13.709&12.900&12.184&0.004&0.004&0.004&0.006&16&17&15&14 \\
C3  &124  &19 49 19.3 &+30 12 49  &3.4  &13.409&12.856&12.524&12.236&0.009&0.008&0.008&0.015&10&10&9&9   \\
C4  &133  &19 49 31.4 &+30 12 35  &0.8  &13.625&13.097&12.772&12.459&0.007&0.008&0.006&0.017&14&14&12&11 \\
C5  &150  &19 49 55.1 &+30 12 16  &4.2  &14.200&13.742&13.451&13.140&0.012&0.009&0.004&0.012&9&9&9&9     \\
C6  &191  &19 49 38.3 &+30 10 56  &1.7  &14.545&13.840&13.414&13.041&0.006&0.006&0.004&0.011&17&18&17&15 \\
C7  &192  &19 49 40.0 &+30 10 55  &1.8  &15.179&14.155&13.501&12.983&0.005&0.006&0.006&0.011&16&18&18&15 \\
C8  &208  &19 49 14.3 &+30 10 27  &5.0  &14.154&13.484&13.074&12.708&0.007&0.007&0.005&0.011&13&13&13&11 \\
C9  &224  &19 49 39.1 &+30 09 26  &3.1  &14.694&13.960&13.496&13.079&0.008&0.007&0.006&0.012&17&17&17&15 \\
C10 &228  &19 49 46.2 &+30 09 17  &3.9  &15.014&14.313&13.873&13.477&0.006&0.006&0.004&0.009&16&18&18&14 \\
\hline
\multicolumn{16}{c}{EXO\,2030+375} \\
\hline
C1  &56  &20 32 10.1 &+37 39 56  &1.9   &16.567&15.586&15.034&14.512&0.009&0.007&0.009&0.009&9&9&9&9    \\
C2  &63  &20 32 22.8 &+37 39 00  &1.6   &18.414&16.949&16.066&15.248&0.016&0.008&0.012&0.025&9&9&9&9    \\
C3  &72  &20 32 27.5 &+37 38 13  &2.4   &17.735&16.462&15.683&14.879&0.021&0.013&0.008&0.011&7&7&7&7    \\
C4  &77  &20 32 34.2 &+37 37 17  &3.3   &17.941&16.634&15.882&15.189&0.012&0.011&0.009&0.011&9&9&9&9    \\
C5  &80  &20 32 26.5 &+37 37 07  &2.5   &18.321&17.058&16.328&15.662&0.012&0.010&0.009&0.010&9&9&9&9    \\
C6  &81  &20 32 21.5 &+37 37 01  &1.7   &18.622&16.746&15.440&14.074&0.010&0.011&0.010&0.013&9&9&9&9    \\
C7  &83  &20 32 07.7 &+37 36 56  &1.9   &18.414&17.044&16.237&15.424&0.012&0.011&0.009&0.009&9&9&9&9    \\
C8  &88  &20 32 16.4 &+37 36 21  &1.9   &17.778&16.636&15.943&15.355&0.010&0.011&0.008&0.026&9&9&9&9    \\
C9  &101 &20 31 55.3 &+37 36 14  &4.3   &17.403&16.102&15.359&14.628&0.012&0.011&0.010&0.010&8&8&8&8   \\
C10 &148 &20 32 14.7 &+37 35 46  &2.5   &17.331&16.108&15.405&14.712&0.017&0.014&0.013&0.013&9&9&9&9   \\
C11 &160 &20 32 18.8 &+37 34 56  &3.4   &17.019&15.467&14.487&13.454&0.014&0.013&0.011&0.016&9&9&8&8  \\
C12 &183 &20 32 14.2 &+37 34 14  &4.1   &17.793&16.510&15.704&15.068&0.014&0.010&0.009&0.015&9&9&9&9  \\
\hline\hline
\end{tabular}
\end{center}
\end{sidewaystable*}

\begin{sidewaystable*}
\caption{Secondary standards in the field of view of high-mass X-ray
binaries (cont.)}
\label{secstd5}
\begin{center}
\begin{tabular}{lcccccccccccccccc}
\hline  \hline

Star &ID &RA    &DEC            &d      &$\overline{B}$ &$\overline{V}$ &$\overline{R}$ &$\overline{I}$ &$\sigma_{\overline B}$     &$\sigma_{\overline V}$  &$\sigma_{\overline R}$ &$\sigma_{\overline I}$ &$N_B$      &$N_V$  &$N_R$  &$N_I$  \\
     &   &      &               &(\arcmin)&(mag)        &(mag)          &(mag)         &(mag)          &(mag)                  &(mag)                  &(mag)                 &(mag)          &        &       &       &       \\     
\hline  \hline                                                  
\multicolumn{16}{c}{GRO\,J2058+42} \\
\hline
C1  &19  &20 59 00.7 &+41 50 06  &4.3  &15.217&14.318&13.786&13.157&0.006&0.006&0.005&0.009&14&14&14&12 \\
C2  &33  &20 58 59.5 &+41 49 02  &3.3  &14.392&13.575&13.101&12.600&0.011&0.005&0.005&0.009&14&14&14&12 \\
C3  &45  &20 58 30.5 &+41 48 14  &3.5  &15.672&14.301&13.482&12.583&0.010&0.004&0.004&0.007&17&17&17&15 \\
C4  &50  &20 58 38.0 &+41 47 52  &2.0  &15.590&14.482&13.832&13.161&0.010&0.004&0.006&0.011&19&19&19&16 \\
C5  &83  &20 59 08.8 &+41 46 24  &3.9  &15.743&14.877&14.377&13.922&0.011&0.007&0.006&0.007&15&15&15&13 \\
C6  &90  &20 58 46.0 &+41 45 57  &0.7  &15.494&14.622&14.092&13.550&0.009&0.006&0.006&0.005&21&21&21&18 \\
C7  &91  &20 58 44.5 &+41 45 51  &0.9  &14.990&14.057&13.511&12.936&0.008&0.005&0.004&0.005&21&21&21&18 \\
C8  &100 &20 59 05.5 &+41 44 20  &4.0  &15.422&14.339&13.711&13.011&0.009&0.006&0.005&0.007&18&18&18&16 \\
C9  &101 &20 59 09.0 &+41 44 05  &4.7  &16.455&15.136&14.352&13.528&0.013&0.006&0.006&0.011&15&15&15&14 \\
\hline  \hline                                                  
\multicolumn{16}{c}{SAX\,J2103.5+4545} \\
\hline
C1  &27  &21 03 35.7 &+45 47 57  &2.8  &15.784&14.772&14.145&13.538&0.010&0.006&0.003&0.012&10&9&10&9    \\
C2  &61  &21 03 51.2 &+45 47 20  &3.4  &15.381&14.664&14.221&13.759&0.013&0.006&0.006&0.009&11&10&11&10  \\
C3  &69  &21 03 10.2 &+45 47 04  &4.8  &16.016&14.875&14.138&13.391&0.011&0.007&0.007&0.011&10&9&10&9    \\
C4  &72  &21 03 36.0 &+45 46 37  &1.4  &15.660&14.465&13.710&12.969&0.007&0.005&0.005&0.008&17&16&17&14  \\
C5  &100 &21 03 36.4 &+45 45 23  &0.3  &14.861&14.080&13.609&13.153&0.006&0.006&0.006&0.010&18&18&19&16  \\
C6  &104 &21 03 20.2 &+45 45 15  &2.7  &15.108&13.585&12.672&11.810&0.008&0.004&0.006&0.007&19&18&12&16  \\
C7  &113 &21 03 17.2 &+45 44 56  &3.2  &16.022&14.772&13.982&13.200&0.006&0.004&0.004&0.007&18&18&19&16  \\
C8  &114 &21 03 22.0 &+45 44 56  &2.4  &15.776&14.690&14.024&13.399&0.006&0.005&0.005&0.006&18&18&19&16  \\
C9  &120 &21 03 27.8 &+45 44 36  &1.4  &15.162&14.302&13.752&13.192&0.009&0.006&0.006&0.010&19&18&19&16  \\
C10 &131 &21 03 29.5 &+45 43 58  &1.5  &15.124&14.190&13.606&13.020&0.007&0.005&0.004&0.005&18&18&19&16  \\
C11 &136 &21 03 10.0 &+45 43 34  &4.6  &15.265&14.348&13.768&13.156&0.008&0.006&0.006&0.009&14&13&14&12  \\
C12 &138 &21 03 30.3 &+45 43 29  &1.8  &15.512&14.697&14.183&13.638&0.007&0.005&0.004&0.007&18&18&19&16  \\
C13 &150 &21 03 42.2 &+45 42 30  &2.7  &14.680&14.166&13.848&13.471&0.007&0.005&0.006&0.007&18&18&19&16  \\
C14 &170 &21 03 52.4 &+45 40 49  &5.1  &15.248&14.424&13.937&13.479&0.007&0.005&0.006&0.009&14&14&15&14  \\
\hline
\multicolumn{16}{c}{IGR\,J21343+4738} \\
\hline
C1  &7   &21 33 55.5 &+47 42 37  &6.2  &15.332&14.446&13.889&13.309&0.006&0.016&0.012&0.011&5&5&5&5      \\
C2  &15  &21 33 55.7 &+47 42 00  &5.7  &14.375&13.607&13.104&12.562&0.006&0.014&0.011&0.012&5&5&5&5      \\
C3  &25  &21 34 09.0 &+47 41 07  &3.6  &14.357&13.632&13.203&12.744&0.010&0.011&0.012&0.010&6&6&5&6      \\
C4  &41  &21 33 57.5 &+47 40 16  &4.4  &14.796&13.958&13.450&12.935&0.007&0.011&0.009&0.009&6&6&6&6      \\
C5  &60  &21 34 30.9 &+47 39 34  &2.3  &15.206&14.356&13.842&13.349&0.008&0.011&0.009&0.010&6&6&6&6      \\
C6  &73  &21 34 38.7 &+47 38 19  &3.0  &14.350&13.683&13.265&12.840&0.008&0.013&0.012&0.012&6&6&5&6      \\
C7  &78  &21 34 30.3 &+47 38 03  &1.6  &15.293&14.432&13.917&13.439&0.009&0.012&0.011&0.010&6&6&6&6      \\
C8  &80  &21 34 35.0 &+47 37 55  &2.4  &14.391&14.014&13.785&13.513&0.010&0.014&0.011&0.014&6&6&6&6      \\
C9  &95  &21 34 45.3 &+47 37 30  &4.1  &14.916&14.418&14.106&13.783&0.008&0.015&0.013&0.017&5&5&5&5      \\
C10 &103 &21 34 12.3 &+47 37 28  &1.3  &15.133&14.157&13.584&13.092&0.008&0.013&0.010&0.010&6&6&6&6      \\
C11 &112 &21 34 24.7 &+47 36 46  &1.4  &14.636&13.865&13.395&12.924&0.009&0.012&0.010&0.012&6&6&6&6      \\
C12 &116 &21 34 44.3 &+47 36 17  &4.4  &14.879&14.113&13.634&13.173&0.007&0.014&0.012&0.015&5&5&5&5      \\
C13 &148 &21 33 59.1 &+47 34 22  &5.0  &14.529&13.763&13.307&12.859&0.009&0.012&0.010&0.011&6&6&6&6      \\
\hline  \hline                                                                                          
\end{tabular}
\end{center}
\end{sidewaystable*}

\begin{sidewaystable*}
\caption{Secondary standards in the field of view of high-mass X-ray
binaries (cont.)}
\label{secstd4}
\begin{center}
\begin{tabular}{lcccccccccccccccc}
\hline  \hline

Star &ID &RA    &DEC            &d      &$\overline{B}$ &$\overline{V}$ &$\overline{R}$ &$\overline{I}$ &$\sigma_{\overline B}$     &$\sigma_{\overline V}$  &$\sigma_{\overline R}$ &$\sigma_{\overline I}$ &$N_B$      &$N_V$  &$N_R$  &$N_I$  \\
     &   &      &               &(\arcmin)&(mag)        &(mag)          &(mag)         &(mag)          &(mag)                  &(mag)                  &(mag)                 &(mag)          &        &       &       &       \\     

\hline  \hline                                                                                          
\multicolumn{16}{c}{4U\,2206+54} \\
\hline
C1  &20  &22 07 59.0 &+54 33 53  &2.8   &13.349&12.902&12.651&12.371&0.006&0.004&0.002&0.007&25&22&23&21 \\
C2  &37  &22 08 20.1 &+54 32 53  &3.9   &14.405&13.766&13.381&12.934&0.007&0.004&0.005&0.005&26&23&24&22 \\
C3  &41  &22 07 30.2 &+54 32 47  &4.0   &14.009&13.633&13.400&13.104&0.007&0.007&0.008&0.008&16&16&16&15 \\
C4  &69  &22 08 05.0 &+54 30 52  &1.3   &14.496&13.753&13.297&12.868&0.009&0.005&0.003&0.005&28&24&25&23 \\
C5  &77  &22 08 20.8 &+54 30 33  &3.5   &10.648&10.429&10.284&10.120&0.017&0.006&0.005&0.006&27&27&27&24 \\
C6  &84  &22 07 48.2 &+54 30 17  &1.4   &13.784&13.132&12.732&12.353&0.006&0.005&0.003&0.005&28&25&24&24 \\
C7  &97  &22 08 25.4 &+54 29 07  &4.6   &12.161&11.950&11.789&11.628&0.007&0.005&0.004&0.007&22&19&20&20 \\
C8  &99  &22 08 15.4 &+54 28 56  &3.4   &14.320&13.806&13.489&13.163&0.006&0.004&0.004&0.006&28&24&26&25 \\
C9  &111 &22 07 59.4 &+54 28 05  &2.9   &13.875&13.266&12.880&12.508&0.007&0.004&0.004&0.005&28&26&26&24 \\
C10 &117 &22 08 01.0 &+54 27 42  &3.4   &12.376&11.835&11.504&11.171&0.006&0.004&0.003&0.006&27&25&27&24 \\
\hline                                                  
\multicolumn{16}{c}{SAX\,J2239.3+6116} \\
\hline
C1  &21  &22 39 16.7 &+61 18 19  &1.9   &17.859&16.425&15.561&14.718&0.009&0.013&0.011&0.007&7&7&7&7 \\
C2  &25  &22 39 12.5 &+61 17 12  &1.2   &15.848&14.774&14.158&13.515&0.011&0.012&0.011&0.009&7&7&7&7 \\
C3  &27  &22 39 14.8 &+61 17 08  &1.0   &18.178&16.433&15.362&14.247&0.008&0.012&0.010&0.008&7&7&7&7 \\
C4  &28  &22 39 32.9 &+61 17 04  &1.5   &18.119&16.778&16.013&15.120&0.010&0.014&0.009&0.008&7&7&7&7 \\
C5  &29  &22 39 34.7 &+61 16 44  &1.7   &17.666&16.351&15.580&14.717&0.008&0.013&0.013&0.013&7&7&7&7 \\
C6  &31  &22 38 54.6 &+61 16 36  &3.1   &16.242&15.142&14.484&13.804&0.010&0.012&0.009&0.008&7&7&7&7 \\
C7  &39  &22 38 53.0 &+61 16 04  &3.3   &17.610&16.252&15.373&14.629&0.012&0.014&0.010&0.008&7&7&7&7 \\
C8  &40  &22 39 38.2 &+61 15 50  &2.2   &16.394&15.249&14.560&13.867&0.013&0.012&0.010&0.007&7&7&7&7 \\
C9  &41  &22 39 30.3 &+61 15 45  &1.3   &16.748&15.656&15.022&14.321&0.011&0.010&0.009&0.008&7&7&7&7 \\
C10 &45  &22 39 35.7 &+61 15 10  &2.1   &16.076&14.974&14.336&13.748&0.010&0.010&0.010&0.007&7&7&7&7 \\
C11 &47  &22 39 48.9 &+61 15 03  &3.6   &18.107&16.495&15.516&14.531&0.009&0.013&0.013&0.007&7&7&7&7 \\
C12 &50  &22 39 43.5 &+61 14 38  &3.2   &17.323&16.028&15.249&14.483&0.011&0.014&0.011&0.010&7&7&7&7 \\
\hline\hline
\end{tabular}
\end{center}
\end{sidewaystable*}

\begin{figure*}
\begin{center}
\begin{tabular}{cc}
\includegraphics[width=7.7cm]{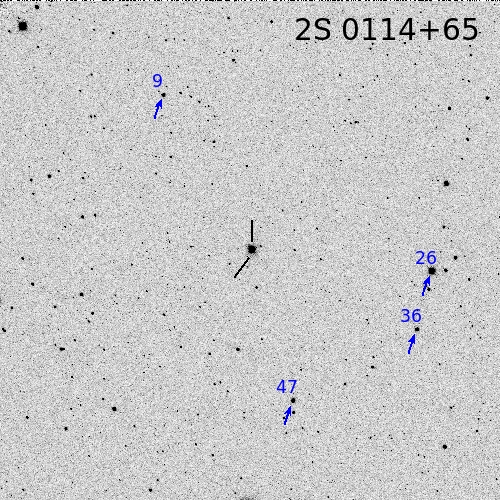} &
\includegraphics[width=7.7cm]{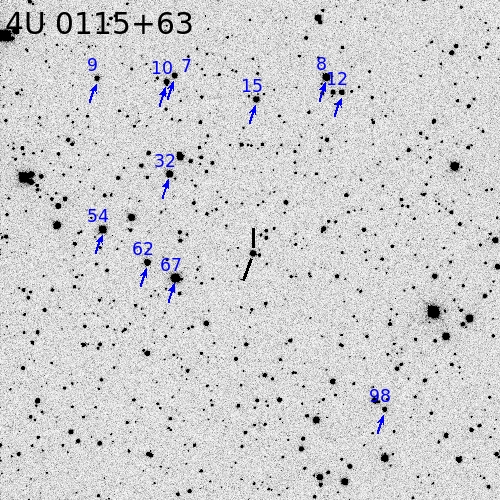} \\
\includegraphics[width=7.7cm]{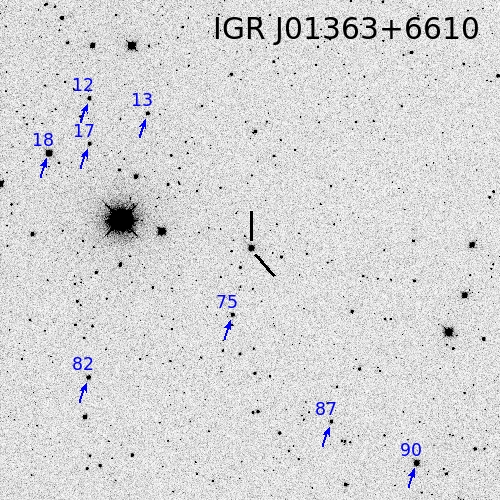} &
\includegraphics[width=7.7cm]{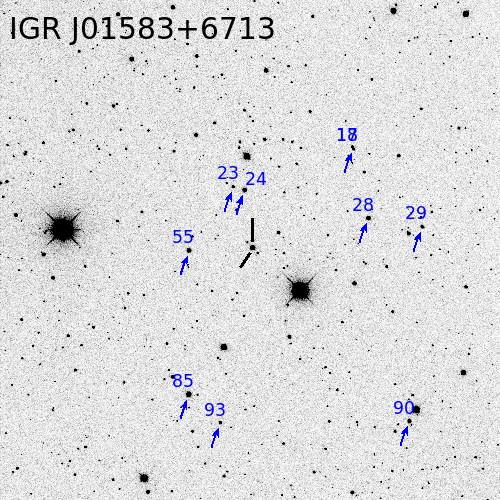} \\
\includegraphics[width=7.7cm]{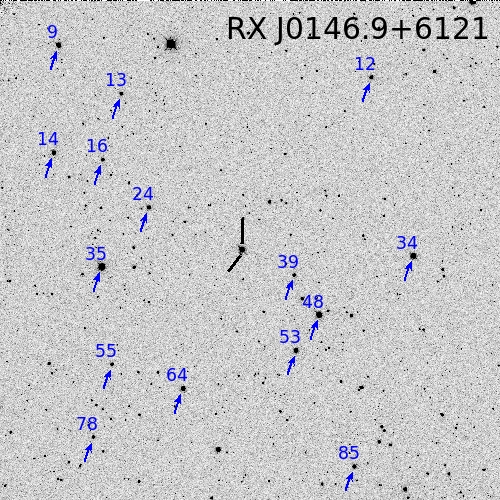} &
\includegraphics[width=7.7cm]{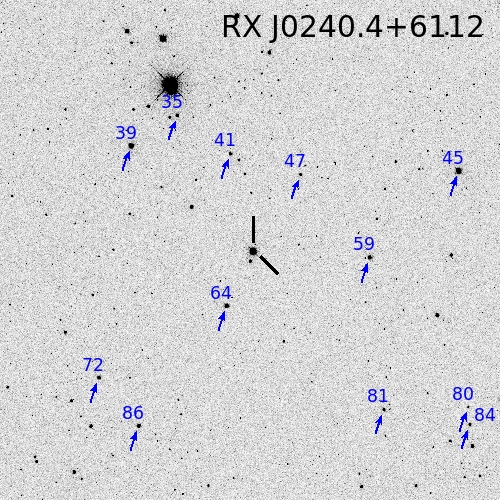} \\
\end{tabular}
\end{center}
\caption[]{Bias-subtracted and flat field-corrected 
R-band images obtained with the 1.3 m telescope of the 
Skinakas observatory. Arrows mark the position of the 
secondary standards. North is up, east is left. 
The field of view is $\sim 9.5' \times 9.5'$. The target 
lies in the centre of the image and it is marked with two lines.}
\label{charts1}
\end{figure*}
\begin{figure*}
\begin{center}
\begin{tabular}{cc}
\includegraphics[width=7.7cm]{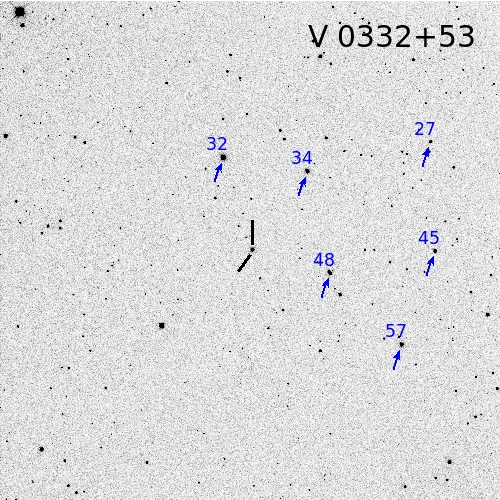} &
\includegraphics[width=7.7cm]{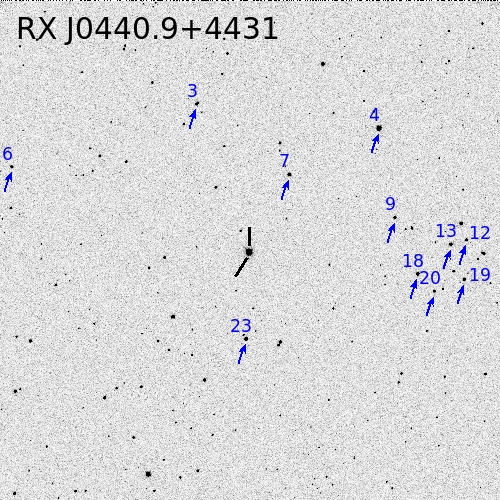} \\
\includegraphics[width=7.7cm]{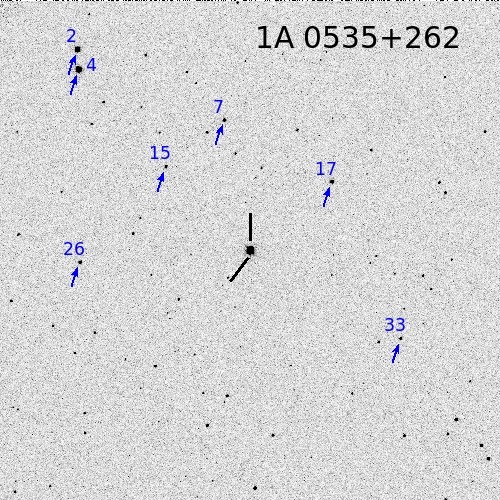} &
\includegraphics[width=7.7cm]{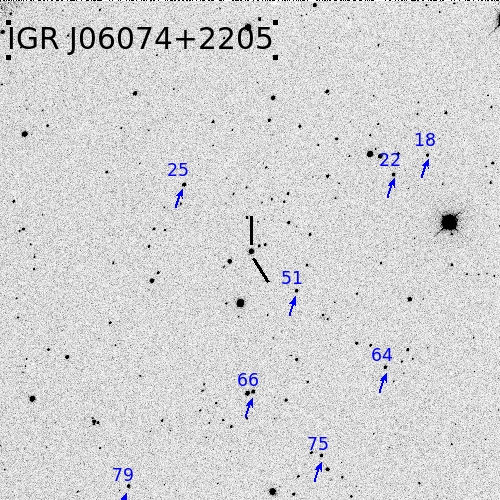} \\
\includegraphics[width=7.7cm]{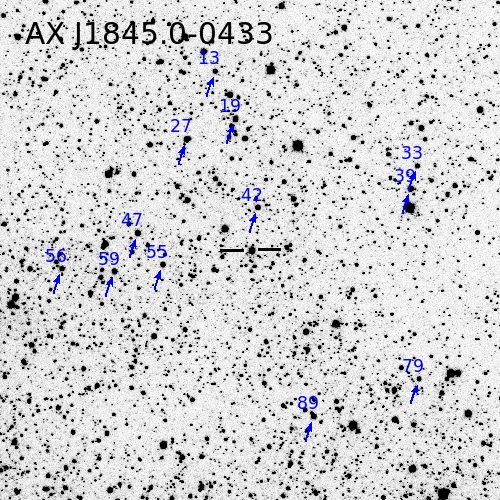} &
\includegraphics[width=7.7cm]{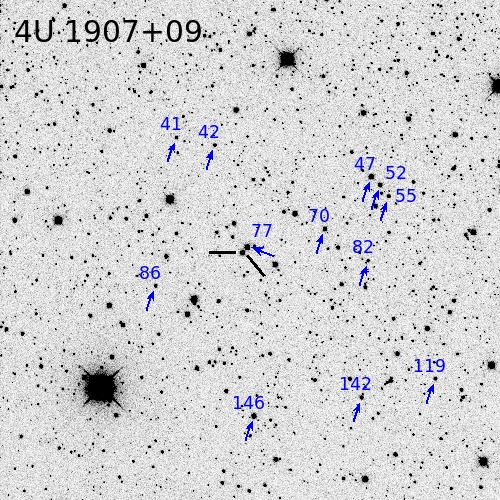} \\
\end{tabular}
\end{center}
\caption[]{Bias-subtracted and flat field-corrected 
R-band images obtained with the 1.3 m telescope of the 
Skinakas observatory. Arrows mark the position of the 
secondary standards. North is up, east is left. 
The field of view is $\sim 9.5' \times 9.5'$. The target 
lies in the centre of the image and it is marked with two lines.}
\label{charts2}
\end{figure*}
\begin{figure*}
\begin{center}
\begin{tabular}{cc}
\includegraphics[width=7.7cm]{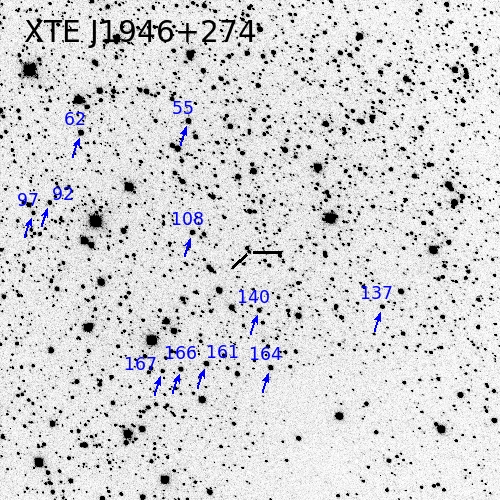} &
\includegraphics[width=7.7cm]{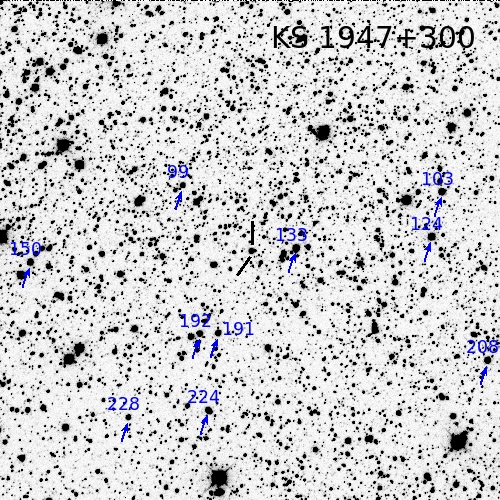} \\
\includegraphics[width=7.7cm]{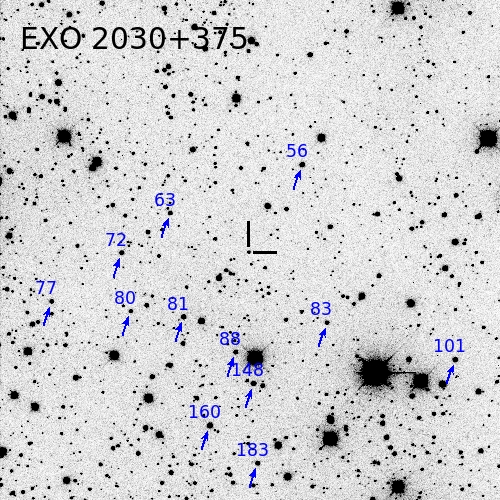} &
\includegraphics[width=7.7cm]{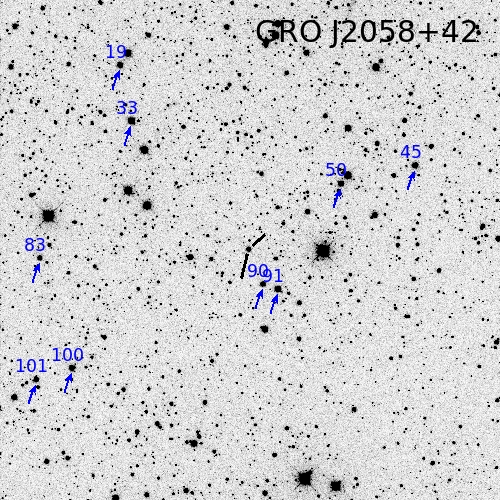} \\
\includegraphics[width=7.7cm]{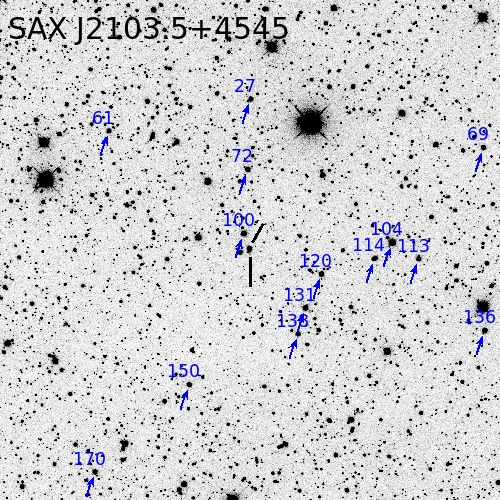} &
\includegraphics[width=7.7cm]{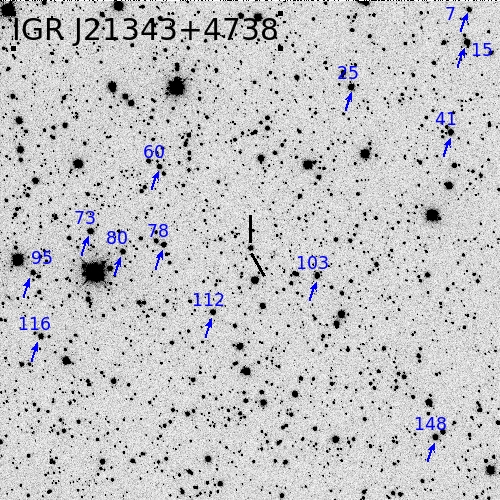} \\
\end{tabular}
\end{center}
\caption[]{Bias-subtracted and flat field-corrected 
R-band images obtained with the 1.3 m telescope of the 
Skinakas observatory. Arrows mark the position of the 
secondary standards. North is up, east is left. 
The field of view is $\sim 9.5' \times 9.5'$. The target 
lies in the centre of the image and it is marked with two lines.}
\label{charts3}
\end{figure*}
\begin{figure*}
\begin{center}
\begin{tabular}{cc}
\includegraphics[width=7.7cm]{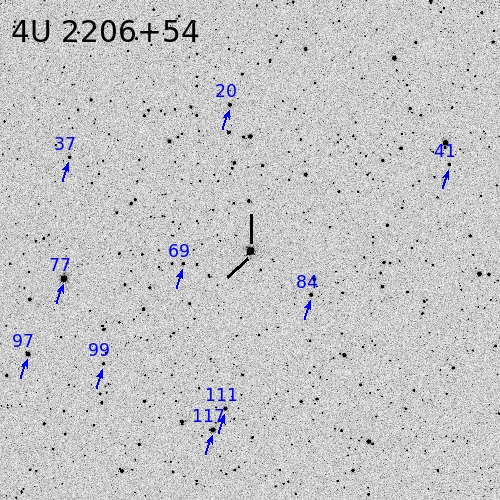} &
\includegraphics[width=7.7cm]{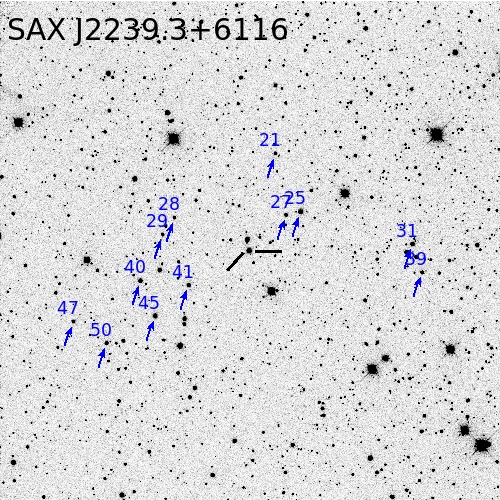} \\
\end{tabular}
\end{center}
\caption[]{Bias-subtracted and flat field-corrected 
R-band images obtained with the 1.3 m telescope of the 
Skinakas observatory. Arrows mark the position of the 
secondary standards. North is up, esast is left. 
The field of view is $\sim 9.5' \times 9.5'$. The target 
lies in the centre of the image and it is marked with two lines.}
\label{charts4}
\end{figure*}

\end{appendix}

\end{document}